\documentclass[%
  reprint,
 amsmath,amssymb,
 aps,
 longbibliography
]{revtex4-2}
\AtBeginDocument{\nocite{apsrev42Control}}
\usepackage{graphicx}
\usepackage{dcolumn}
\usepackage{bm}
\usepackage{xcolor}
\usepackage[caption=false]{subfig}
\usepackage{overpic}
\usepackage{makecell}
\usepackage{multirow}
\usepackage{hyperref}
\usepackage{booktabs}
\usepackage{threeparttable}
\usepackage{array}

\begin{document}
\nocite{apsrev42Control}

\preprint{APS/123-QED}

\title{Qubit-Efficient Variational Quantum Optimization via Pauli Correlation Encoding: \\
Application to Large-Scale Power Demand Portfolio Optimization}

\author{Takuya Yoshioka}
\email{takuya.yoshioka@tisi.jp}
\affiliation{Strategic Technology Center, TISI Inc., 2-2-1 Toyosu, Koto-ku, Tokyo 135-0061, Japan}

\author{Keita Sasada}
\affiliation{Strategic Technology Center, TISI Inc., 2-2-1 Toyosu, Koto-ku, Tokyo 135-0061, Japan}

\author{Riku Usuki}
\affiliation{Graduate School of Engineering Science, Osaka University, 
1-3 Machikaneyama, Toyonaka, Osaka 560-8531, Japan}

\author{Yuichiro Nakano}
\affiliation{Graduate School of Engineering Science, Osaka University, 
1-3 Machikaneyama, Toyonaka, Osaka 560-8531, Japan}

\author{Keisuke Fujii}
\affiliation{Graduate School of Engineering Science, Osaka University, 
1-3 Machikaneyama, Toyonaka, Osaka 560-8531, Japan}
\affiliation{Center for Quantum Information and Quantum Biology, 
Osaka University, 1-2 Machikaneyama, Toyonaka, Osaka 560-0043, Japan}
\affiliation{Center for Quantum Computing, RIKEN, 
2-1 Hirosawa, Wako, Saitama 351-0198, Japan}





\date{\today}

\begin{abstract}
Variational quantum algorithms offer a promising route to combinatorial optimization, but their applicability is limited by the challenge of encoding large-scale problems within restricted qubit resources.
In this work, we introduce a qubit-efficient variational framework based on Pauli correlation encoding (PCE) and apply it to electric power demand portfolio optimization.
Binary variables are represented through expectation values of Pauli correlation operators, which encode multi-body correlations of the quantum state and provide a continuous relaxation enabling compact representations with few qubits.
We further propose a two-stage hybrid formulation, in which a time-averaged problem provides initialization for a time-resolved optimization.
Numerical simulations demonstrate near-optimal performance across problem sizes ranging from $m=18$ to $10{,}296$, with normalized cost gaps on the order of $10^{-4}$ relative to solutions with certified optimality.
We show that the performance is governed by the interplay between continuous relaxation and discretization:
the effective resolution of the correlator representation determines how reliably improvements in the continuous loss translate into better discrete solutions, with larger systems exhibiting more consistent behavior.
Finally, we demonstrate robustness on a trapped-ion quantum processor, where high-quality solutions are obtained despite noise and finite sampling.
These results establish PCE as a physically motivated and qubit-efficient framework for large-scale combinatorial optimization.
  \end{abstract}

\maketitle


\section{Introduction}
Variational quantum algorithms have emerged
as a flexible framework
for tackling combinatorial optimization problems
on near-term quantum devices~\cite{Peruzzo2014,Farhi2014QAOA,Biamonte2017,Mitarai2018,Cerezo2021}.
However, they face a fundamental scalability challenge:
current hardware provides only a limited number of qubits,
whereas many practically relevant optimization problems
involve extremely large classical variable spaces
together with dense quadratic interactions
and real-valued, nonuniform coefficients.
Addressing this gap requires reconciling
the discrete, high-dimensional structure of classical optimization problems
with the continuous and resource-constrained nature
of quantum representations.
Understanding and controlling this interplay
between discrete variables and their continuous embeddings
remains a central challenge
in variational quantum optimization.

In standard approaches such as the Quantum Approximate Optimization Algorithm (QAOA)~\cite{Farhi2014QAOA,Hadfield2019},
binary optimization variables are typically mapped directly onto qubits,
so the required qubit count grows linearly with problem size.
This becomes a major limitation for large, fully connected quadratic optimization problems,
motivating alternative encoding strategies
that compress the problem representation
while retaining useful optimization structure.

Several approaches have been proposed
to address the qubit-scaling bottleneck,
including space-efficient encodings~\cite{Tan2021,Glos2022}
and quantum random access optimization (QRAO)
with relaxation-based extensions~\cite{Teramoto2023,He2025}.
These methods reduce qubit requirements
by mapping binary optimization problems
to compressed or relaxed quantum representations
followed by measurement and decoding.
However, such compression typically introduces trade-offs
in decoding sensitivity, representational flexibility,
and measurement overhead.
Accordingly, balancing qubit efficiency,
continuous relaxation, and sampling cost
remains an open challenge
for large-scale combinatorial optimization~\cite{Tan2021,Glos2022,Teramoto2023,Sharma2024,Matsuyama2024,He2025}.

These limitations motivate the exploration of alternative representations  
based on physically meaningful observables of quantum states.  
In particular, expectation values of Pauli operators  
encode multi-body correlations of the underlying quantum state  
and provide a continuous description of discrete degrees of freedom.  
This perspective connects combinatorial optimization  
to the statistical structure of quantum states.  
Leveraging this perspective,
Pauli correlation encoding (PCE),
introduced by Sciorilli \textit{et al.}~\cite{Sciorilli2025PCE},
represents classical binary variables
not as individual qubits,
but as the signs of expectation values of Pauli correlation operators
generated by a variational quantum circuit.

\begin{figure}[htb]
  \centering
    \includegraphics[width=7.9cm]{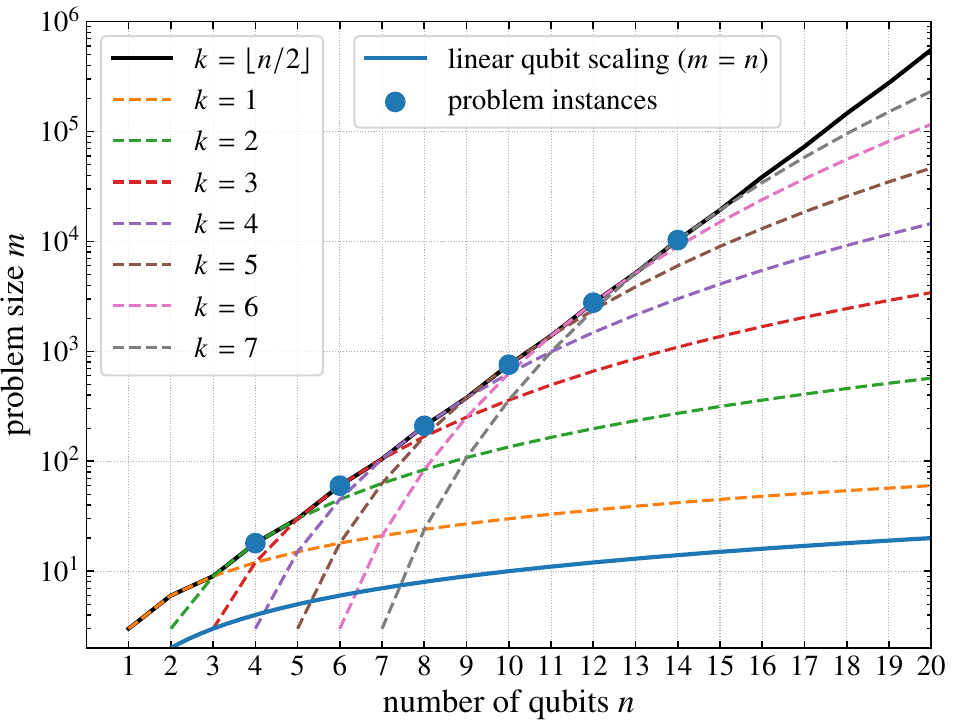}
    \caption{\label{fig:curve_compression}
Maximum representable problem size $m$
as a function of the number of qubits $n$
for different correlation orders $k$
in Pauli correlation encoding (PCE)~\cite{Sciorilli2025PCE},
where $m = 3\binom{n}{k}$.
The black solid curve corresponds to
$k=\lfloor n/2 \rfloor$,
which maximizes the number of Pauli correlators
(i.e., expectation values of Pauli correlation operators).
The linear reference line ($m=n$)
indicates direct encodings with one-to-one qubit--variable mapping.
Filled markers denote the $(n,m)$ instances considered in this work.
}
\end{figure}

For a fixed correlation order $k$,
PCE represents classical variables
using the expectation values
$\{\langle \Pi_i^{(k)} \rangle\}_{i=1}^{m}$
of $k$-body Pauli correlation operators.
These expectation values, hereafter referred to as Pauli correlators,
can be interpreted as continuous relaxations of binary variables
that retain information about higher-order correlations in the quantum state.
Their number scales combinatorially as
$m = 3\binom{n}{k}$,
enabling the representation of a large number of classical variables
with comparatively few qubits.
In particular, choosing
$k=\lfloor n/2 \rfloor$
maximizes the correlator count,
as shown in Fig.~\ref{fig:curve_compression}.

Throughout this study, we focus on even $n$ and adopt $k=n/2$,
which maximizes the number of correlators for a fixed $n$.
Importantly, this large number of available correlators does not imply
an equivalent increase in the number of independent degrees of freedom.
All correlators are generated from the same underlying quantum state
and are therefore constrained by quantum correlations and the expressivity
of the variational ansatz.
Accordingly, PCE should be viewed as a compressed representation
rather than an exact embedding of independently controllable variables.
The present work does not assume independent control of all encoded variables,
but instead investigates whether useful optimization structure can be represented
and exploited within the constrained manifold generated by the variational quantum circuit.

Following the original proposal~\cite{Sciorilli2025PCE},
PCE has been studied
for several combinatorial optimization problems,
including large-scale benchmark instances such as Max-Cut~\cite{Sciorilli2025PCE},
smaller-scale LABS problems on quantum hardware~\cite{Sciorilli2025LABS},
and comparative simulation studies across multiple problem classes~\cite{Sharma2025Comparative}.
Applications to more practical tasks, including portfolio optimization,
have also been explored,
but mainly at moderate scales and in simulation~\cite{Soloviev2026Portfolio,Padinmartinez2026}.
However, the behavior of PCE remains insufficiently understood
for dense quadratic problems
with real-valued and nonuniform coefficients.

In this work, we study PCE in this regime
through electric power demand portfolio optimization,
which naturally yields fully connected QUBO instances.
This application provides a useful setting
for examining the scalability and optimization behavior of PCE
beyond structured benchmark problems.
A central issue is not only whether the relaxed loss can be minimized,
but whether such improvements are reliably transmitted
to the discretized binary solution.
This motivates a systematic investigation
of the relationship among continuous relaxation,
binary decoding, and the distribution of Pauli correlators.

The novelty of the present work does not primarily lie in
applying PCE to another optimization problem,
but rather in clarifying how a compact correlator-based
representation behaves in dense quadratic optimization problems
with real-valued and nonuniform coefficients.
Specifically, this work provides three contributions.

First, we investigate the behavior of PCE in a class of
fully connected quadratic optimization problems with
real-valued and nonuniform coefficients,
using electric power demand portfolio optimization
as a representative application.
This setting differs substantially from the benchmark problems
primarily considered in previous PCE studies and provides a useful
testbed for examining the representational and optimization
properties of PCE in practically motivated dense QUBO instances.
We demonstrate near-optimal performance,
with normalized cost gaps on the order of $10^{-4}$
relative to solutions with certified optimality
obtained using a state-of-the-art commercial solver~\cite{gurobi},
across a wide range of problem sizes.
Second, we show that the optimization behavior of PCE
is governed by the interplay between the continuously relaxed loss
and the discretized binary objective.
To analyze this mechanism, we study the distribution of Pauli correlators,
which serve as continuous representations of binary variables,
and characterize how this interplay between continuous relaxation
and discrete binary structure manifests in the correlator space.
As a result, we clarify how the effective resolution of the correlator representation
evolves with system size and influences the optimization behavior.
Third, we validate the robustness of this representation
on trapped-ion quantum hardware
and show that useful binary solutions can still be recovered
through post-processing
under finite sampling and hardware noise.
This result provides practical evidence that the continuous relaxation framework
retains sufficient information about the underlying binary structure,
enabling reliable solution recovery even in the presence of hardware imperfections.

These contributions provide new insight into the representational capacity,
optimization behavior, and practical robustness of PCE
beyond previously studied benchmark settings.
Taken together, these results provide a unified understanding of PCE
from both representational and optimization perspectives
in fully connected settings with practical relevance.
In particular, our findings clarify how continuous relaxation
can faithfully encode binary structures,
how the interplay between continuous and discrete representations
governs the optimization mechanism,
and how these properties persist under realistic hardware conditions.
This positions PCE as a viable framework for bridging
compact quantum representations and practical combinatorial optimization,
offering a complementary approach to existing quantum and classical methods
in the NISQ regime.

This paper is organized as follows.
Section~\ref{sec:power_portfolio}
introduces the power demand portfolio optimization problem
and its time-averaged (Model~1)
and time-resolved (Model~2) formulations.
Section~\ref{sec:pce_framework}
presents the PCE-based variational framework.
Section~\ref{sec:numerical_results}
reports the numerical and hardware results.
Sections~\ref{sec:discussion}
and~\ref{sec:conclusion}
provide discussion and conclusion.
Additional analyses and supplementary results
are provided in the Appendix.

\section{Electric Power Demand Portfolio Optimization Problem}
\label{sec:power_portfolio}

Electric power demand portfolio optimization is a practically important task
in power aggregator operations,
where subsets of consumers are selected to meet procurement targets
under stochastic uncertainty in available demand reduction~\cite{Albadi2008,Palensky2011}.
The resulting problem is typically large-scale and combinatorial,
with dense quadratic couplings arising from correlations among consumer demand reduction profiles
~\cite{CRIEPI_C18005,Faia2021,Ajagekar2019,YoshiokaFQAOA2024, YoshiokaQCE2026}.
This structure makes it a natural benchmark for high-dimensional QUBO-type optimization
with real-valued, nonuniform coefficients.

\subsection{Problem Definition and Data Construction}
\label{subsec:problem_setting}

We consider a power demand portfolio optimization problem
in which a power aggregator selects a subset of demand-side consumers
to procure a target amount of flexible power under uncertainty~\cite{CRIEPI_C18005,YoshiokaFQAOA2024}.
Let $p_{t,i}$ denote the available demand reduction
that can be procured from consumer $i$ at time $t$,
treated as a random variable reflecting stochastic variability in available demand reduction.
The portfolio selection is represented by binary variables
${\bm x}=(x_1,\ldots,x_m)\in\{0,1\}^m$,
where $x_i=1$ indicates that consumer $i$ is included in the portfolio
and $x_i=0$ otherwise.

The model parameters are derived from open electricity consumption data
provided by the Energy Management System Open Data platform~\cite{ems_opendata}.
The dataset consists of hourly electricity consumption records
covering a 61-day period from April~1 to May~30,
and includes $2{,}143$ independent consumer-level time series.
For each consumer $i$ and time step $t$,
the available demand reduction $p_{t,i}$ is defined as 10\%
of the original electricity consumption.

For problem instances with $m \le 2{,}143$,
the portfolio is constructed by randomly selecting
$m$ distinct consumers from the available dataset.
For larger-scale instances with $m > 2{,}143$,
additional consumer profiles are generated
using the mixup data augmentation method,
in which synthetic reduction profiles are constructed
as convex combinations of pairs
of independent consumer time series.
This combined procedure enables a systematic study
of problem instances across a wide range of portfolio sizes
while preserving the statistical characteristics
of the original data.

The problem sizes considered in this work
are selected to span representative points
along the scaling curve enabled by Pauli correlation encoding,
as shown in Fig.~\ref{fig:curve_compression},
ranging from small instances to large instances approaching the representational limit
of the available qubit resources.

\subsection{Cost Models: Averaged and Time-Resolved Formulations}
\label{subsec:cost_models}

We introduce two related formulations
of the demand portfolio optimization problem:
a time-averaged model (Model~1)
and a time-resolved model (Model~2).
Both are based on the same hourly procurement cost,
while differing in how temporal fluctuations are treated.

We first define the hourly cost at time step $t$
for a given portfolio configuration ${\bm x}$ as
\begin{align}
  C_t({\bm x})
  =&
  \mathbb{E}\!\left[
    \left(
      \sum_{i=1}^{m} p_{t,i} x_i - P^{\mathrm{target}}_t
    \right)^2
  \right] \nonumber\\
  =&
  \sum_{i=1}^{m}\sum_{j=1}^{m}
  \mathrm{Cov}(p_{t,i},p_{t,j}) x_i x_j \nonumber\\
  &+
    \left(
    \sum_{i=1}^{m}\mathbb{E}[p_{t,i}] x_i
    - P^{\mathrm{target}}_t
  \right)^2 .
\label{eq:hourly_cost}
\end{align}
This objective measures the deviation
between procured power and the procurement target
at each time step,
while incorporating both stochastic fluctuations
and the mean available demand reduction.

Throughout this work,
the target power is set as
\begin{align}
P^{\mathrm{target}}_t
=
\frac{1}{2}\sum_{i=1}^{m} \mathbb{E}[p_{t,i}],
\label{eq:target}
\end{align}
so that the task corresponds to procuring approximately half of the total available demand reduction.
This provides a balanced benchmark setting
while preserving the nontrivial correlation structure
of the procurement problem.

\subsubsection{Model 1: Time-Averaged Portfolio Optimization}
\label{par:model1}

Model~1 is a time-averaged formulation
in which a common portfolio is selected
over a time window of length $n_T$.
This model is introduced
to capture the dominant statistical structure
of uncertainty in available demand reduction
and to provide effective initial parameters
for the subsequent time-resolved optimization.

The averaged cost function is defined as
\begin{align}
  C_T({\bm x})
  &=
  \frac{1}{n_T}
  \sum_{t=T}^{T+n_T-1}
  C_t({\bm x}). \label{eq:total_cost}
\end{align}
Substituting Eq.~(\ref{eq:hourly_cost}) into Eq.~(\ref{eq:total_cost}) gives
\begin{align}
  C_T({\bm x})
  &=
  \sum_{i=1}^{m}\sum_{j=1}^{m}
  \overline{\mathrm{Cov}(p_{t,i},p_{t,j})}
  x_i x_j
  \nonumber \\
  &\quad
  + \frac{1}{n_T}
  \sum_{t=T}^{T+n_T-1}
  \left(
    \sum_{i=1}^{m}\mathbb{E}[p_{t,i}] x_i
    - P^{\mathrm{target}}_t
  \right)^2,
\label{eq:averaged_cost}
\end{align}
where the overline denotes averaging
over the time window of length $n_T$.
In this work,
the reference time is set to $T=1$,
and the averaging window is fixed to $n_T=24$.

For later normalization and comparison of optimization results,
we define reference minimum and maximum cost values,
$C_T^{\min}$ and $C_T^{\max}$,
which are obtained using the Gurobi optimizer~\cite{gurobi}
with a relative optimality gap tolerance of $10^{-4}$.
Using these reference values,
we introduce the normalized cost gap
\begin{equation}
\frac{\Delta C_T}{W_T}
= \frac{C_T - C_T^{\min}}{C_T^{\max} - C_T^{\min}},
\label{eq:normal_cost}
\end{equation}
which takes values between 0 and 1
and quantifies the relative optimality of a given solution.
Similarly, we define the normalized loss gap as
$\Delta L_T/W_T$,
where $\Delta L_T = L_T - C_T^{\min}$ uses the same optimal cost reference,
so that both cost and loss are measured
relative to the same optimal reference.
These normalized gap measures are used throughout this work
to compare optimization performance for Model~1
across different problem sizes and solution methods.

\subsubsection{Model 2: Time-Resolved Portfolio Optimization}
\label{par:model2}

Model~2 corresponds to the fully time-resolved formulation,
in which the hourly cost $C_t({\bm x})$ defined in Eq.~(\ref{eq:hourly_cost})
is minimized independently at each time step $t$.
This formulation represents the practically motivated operational model
used to evaluate procurement performance
under realistic fluctuations in available demand reduction.

For later analysis,
we define the expected procured power
and its variability at each time step $t$.
The expected procured power is given by
\begin{align}
  P_t(\bm{x}) = \sum_{i=1}^{m} \mathbb{E}[p_{t,i}] x_i ,
\label{eq:Pt}
\end{align}
and the corresponding standard deviation is
\begin{align}
\sigma_t(\bm{x}) =
\sqrt{
\sum_{i=1}^{m}\sum_{j=1}^{m}
\mathrm{Cov}(p_{t,i},p_{t,j}) x_i x_j
}.
\label{eq:standard_deviation}
\end{align}

Using these definitions,
the hourly cost in Eq.~(\ref{eq:hourly_cost})
can be written in the compact form
\begin{equation}
C_t(\bm{x}) =
\sigma_t(\bm{x})^2 + \left[
P_t(\bm{x}) - P^{\mathrm{target}}_t
\right]^2.
\label{eq:cost_decomposition}
\end{equation}
These quantities characterize
the mean and variability of the procured power
and are used in Sec.~\ref{sec:numerical_results}
to evaluate portfolio quality and robustness.

In this formulation,
the requirement to match the procurement target
$P_t^{\mathrm{target}}$
acts as a soft constraint through the quadratic penalty term
in Eq.~(\ref{eq:cost_decomposition}).
While such penalty-based formulations are standard
in classical combinatorial optimization,
their behavior becomes nontrivial
when the binary variables are optimized
through a continuous quantum representation.
In particular,
within the PCE framework introduced below,
the optimization is performed
in a continuously relaxed space,
whereas the final portfolio is obtained only after binary decoding.
This interplay between continuous relaxation
and discrete binary structure
underlies the optimization mechanism analyzed later in this work.

In practice, this mechanism is realized through three stages:
(i) variational optimization in the continuous correlator space,
(ii) binary decoding of the resulting expectation values,
and (iii) greedy post-processing to locally refine the decoded solution.
This multi-stage procedure should be distinguished from the greedy (random) baseline,
which applies only step (iii) starting from randomly generated bit strings.

\section{Pauli Correlation Encoding and Variational Framework}
\label{sec:pce_framework}

In this section, we summarize the Pauli correlation encoding (PCE)
used to represent large-scale binary optimization problems
and describe the variational quantum framework
used to optimize the encoded variables.
We first introduce the structure of the PCE representation
and then explain how it is incorporated
into a variational quantum optimization procedure.

\subsection{Pauli Correlation Encoding of the Optimization Problem}
\label{subsec:pce_representation}

Pauli correlation encoding (PCE)~\cite{Sciorilli2025PCE}
provides a qubit-efficient representation
of large-scale combinatorial optimization problems
by embedding many variables
into correlated expectation values of Pauli operators.
As illustrated in Fig.~\ref{fig:curve_compression},
this encoding enables a combinatorial growth
of the effective problem size $m$
with respect to the number of qubits $n$,
allowing realistically sized power demand portfolio problems
to be addressed using a relatively small quantum system.

For a fixed correlation order $k$,
the PCE representation is constructed
from a set of $k$-body Pauli correlation operators
$\{\Pi_i^{(k)}\}_{i=1}^{m}$, where
\begin{equation}
\begin{aligned}
\Pi_i^{(k)}
&\in
\left\{
\prod_{j\in S} X_j,\;
\prod_{j\in S} Y_j,\;
\prod_{j\in S} Z_j
\right\},\\
&\text{with}\quad
S \subset \{1,\dots,n\},
\quad
|S|=k.
\end{aligned}
\label{eq:pce_operator_set}
\end{equation}
Here, $S$ specifies the subset of qubits
on which the operator acts nontrivially.
Each operator is therefore defined
by choosing a subset $S$ of $k$ qubits
and assigning a common Pauli type
($X$, $Y$, or $Z$) to that subset.
The index $i=1,\dots,m$ labels the resulting distinct operators,
whose expectation values
$\{\langle \Pi_i^{(k)} \rangle\}_{i=1}^{m}$
serve as the encoded continuous variables,
following the original PCE construction~\cite{Sciorilli2025PCE}.

Since there are $\binom{n}{k}$ possible subsets
and three possible Pauli types,
the total number of available correlators is
\begin{equation}
m = 3\binom{n}{k}.
\label{eq:num_correlators}
\end{equation}
Importantly, despite the combinatorial scaling of $m$,
the number of distinct measurement settings
does not scale with the number of encoded variables.
Because the operators in Eq.~\eqref{eq:pce_operator_set}
are grouped according to their Pauli type,
all correlators can be estimated from measurements
in only three global bases,
namely the all-$X$, all-$Y$, and all-$Z$ bases.
Consequently,
the number of quantum state preparations
and measurement configurations
remains fixed at three,
independent of the problem size $m$.
This property substantially reduces the measurement-setting overhead
associated with estimating a combinatorially large number
of Pauli correlators.
However, as discussed in Sec.~\ref{sec:discussion},
overall computational cost also depends on
the statistical precision required for correlator estimation,
the cost of evaluating the dense QUBO objective,
and the cost of classical post-processing.

This combinatorial scaling enables the representation
of a large number of classical variables
using a comparatively small number of qubits.
In particular, for even $n$,
the choice $k=n/2$
maximizes the number of distinct Pauli correlation operators
and therefore yields the largest number of available encoded variables
for a fixed number of qubits.
Throughout this study, we adopt this choice.

\subsection{Variational Optimization Framework}
\label{subsec:variational_framework}

To solve the encoded optimization problem,
we employ a variational quantum framework
in which a parameterized quantum circuit generates
the Pauli correlators
$\{\langle \Pi_i^{(k)} \rangle_{\bm \theta}\}_{i=1}^{m}$,
which are then optimized through a classical outer loop.

\subsubsection{Parameterized Quantum Circuit}
\label{subsubsec:ansatz}

In the present work,
the quantum circuit is used
as a learning-based parametric model
that generates Pauli correlators,
rather than as a standalone quantum algorithm.
The parameterized quantum state
$|\psi({\bm \theta})\rangle$
is prepared by a layered circuit ansatz.
Each layer consists of the sequential application of
(i) $R_y$ rotations on all qubits,
(ii) $R_z$ rotations on all qubits, and
(iii) fully connected $R_{zz}$ rotations
between all qubit pairs.
For a circuit with $n$ qubits and depth $l$,
each layer therefore contains
$\binom{n}{2}$ two-qubit $R_{zz}$ rotations
together with $2n$ single-qubit rotations,
corresponding to $\binom{n}{2}+2n$ elementary gates per layer.
As a result,
the total number of entangling gates scales as $O(l\,n^2)$.

Throughout this work,
the circuit depth is fixed to $l=5$.
This choice reflects a trade-off
between expressivity and computational cost:
because each layer already contains
$O(n^2)$ entangling gates,
increasing the depth substantially
would significantly amplify
the total computational cost,
particularly for large-scale instances.

Empirically,
a depth of five was found to provide
stable optimization performance
across all problem sizes considered in this work.
This depth also remains compatible
with both state-vector simulation
and execution on trapped-ion quantum hardware.
A systematic investigation
of the expressibility and depth scaling
of the ansatz is beyond the scope
of the present study.
For each encoded operator $\Pi_i^{(k)}$,
the corresponding correlator is defined as
\begin{equation}
  \langle\Pi_i^{(k)}\rangle_{\bm \theta}
  =
  \langle\psi({\bm\theta})|
  \Pi_i^{(k)}
  |\psi({\bm\theta})\rangle.
\end{equation}
These expectation values
serve as the fundamental continuous quantities
used in the relaxation described below.

\subsubsection{Continuous Relaxation and Loss Function}
\label{subsubsec:loss_function}

To enable variational quantum optimization,
the binary variables ${\bm x}$ are relaxed
to continuous variables
${\bm y}({\bm\theta})\in[0,1]^m$
constructed from the Pauli correlators
generated by the parameterized circuit.
Specifically, each relaxed variable is defined as
\begin{equation}
  y_i({\bm\theta})
  =
  \varsigma\!\left(
    2\alpha
    \langle\Pi_i^{(k)}\rangle_{\bm \theta}
  \right),
\label{eq:relaxed_variable_definition}
\end{equation}
where $\varsigma(x)=[1+e^{-x}]^{-1}=[\tanh(x/2)+1]/2$
is the sigmoid function,
and $\alpha$ controls the sharpness of the relaxation.

This relaxation is central to the present formulation:
the optimization is performed
in the continuous space of ${\bm y}({\bm \theta})$,
whereas the final binary configurations
are evaluated only after binary decoding.
As shown later,
the consistency between these two levels
plays a central role
in the observed optimization behavior.

Using these relaxed variables,
the loss function for the time-resolved model is defined as
\begin{align}
  L_t({\bm\theta})
  &=
  C_t({\bm y}({\bm\theta}))
  +
  \frac{\beta\nu}{4m}
  \sum_{i=1}^{m}
  \left[
    y_i({\bm\theta})-\frac{1}{2}
    \right]^2,
  \label{eq:loss}
\end{align}
where $\beta$ is a tunable regularization parameter
and $\nu$ is the Frobenius norm of the QUBO matrix.
An analogous loss function is defined
for the averaged model
by replacing $C_t$ with $C_T$.
The factor $1/4$ ensures consistency with the equivalent formulation based on $\tanh$
(see Eq.~\eqref{eq:relaxed_variable_definition}), matching the form used in prior work \cite{Sciorilli2025LABS}.
In evaluating $\nu$,
constant terms in the quadratic cost function,
such as ${(P^{\mathrm{target}}_t})^2$,
which do not depend on the variables,
are excluded.

All variational optimizations are performed
using noiseless state-vector simulation
implemented with the Qulacs simulator~\cite{qulacs}.
The variational parameters ${\bm\theta}$
are optimized using the
Broyden--Fletcher--Goldfarb--Shanno (BFGS) algorithm,
which was found to provide stable and efficient convergence
across all problem sizes considered.
Gradients of the loss function
are computed from exact expectation values
using the backpropagation functionality
available in Qulacs~\cite{qulacs}.
For each Model~1 instance,
the variational optimization is repeated
over five independent runs
with different random initial circuit parameters.
Among these runs, we select the one
that yields the minimum decoded cost
with respect to the original binary objective,
and denote the corresponding circuit parameters by
${\bm \theta}^*$.

The resulting parameters ${\bm \theta}^*$
are used both to initialize the optimization for the time-resolved Model~2
and to evaluate the corresponding QPU results,
without introducing additional random restarts.
This warm-start strategy improves convergence speed
and stabilizes the optimization for large-scale instances.
Quantum hardware experiments are performed
by executing the optimized circuits with fixed parameters
on a quantum processing unit,
followed by measurement of the corresponding expectation values
and cost functions.

\subsubsection{Decoding and Post-Processing}
\label{subsubsec:decoding_postprocessing}

After optimization,
binary portfolio selections are obtained
by threshold-based decoding of the correlators:
\begin{align}
  x_i({\bm \theta})
  :=
  \frac{1}{2}
  \left[
    \mathrm{sign}
    \left(
      \langle\Pi_i^{(k)}\rangle_{\bm \theta}
    \right)
    +1
  \right].
\end{align}
In realistic quantum hardware,
the estimation of expectation values
$\langle\Pi_i^{(k)}\rangle_{\bm \theta}$
is fundamentally limited by finite-shot projection noise.
For a finite number of measurement shots
$N_{\mathrm{shot}}$,
the statistical uncertainty of each estimated correlator
scales as $O(1/\sqrt{N_{\mathrm{shot}}})$.
Consequently,
if the true expectation value lies close to the decoding boundary,
i.e.,
$|\langle\Pi_i^{(k)}\rangle_{\bm \theta}|
\lesssim 1/\sqrt{N_{\mathrm{shot}}}$,
its decoded sign becomes increasingly sensitive
to statistical fluctuations,
leading to sign errors in the decoded binary variables.
This sensitivity is particularly relevant in large-scale instances,
where a significant fraction of correlators may be concentrated near zero.
The resulting degradation of the decoded solution quality
is discussed further in Sec.~\ref{sec:hardware_validation}.
The decoded configurations ${\bm x}({\bm \theta})$
are then evaluated using the original binary cost functions,
such as $C_T({\bm x}({\bm \theta}))$ for Model~1
and $C_t({\bm x}({\bm \theta}))$ for Model~2.
This threshold-based decoding
introduces a nonlinear projection
from the relaxed representation
to the binary feasible space,
and therefore plays a central role
in the interplay between continuous relaxation
and discrete binary structure
analyzed in Sec.~\ref{sec:optimization_mechanism}.

Because the decoding map is nonlinear,
the parameter values that minimize the relaxed loss
do not necessarily coincide with those
that minimize the decoded binary cost.
Accordingly, for each optimization run,
we denote by ${\bm \theta}^*$
the variational parameters along the optimization trajectory
that minimize the decoded binary cost
without post-processing,
namely $C_T({\bm x}({\bm \theta}))$ for Model~1
or $C_t({\bm x}({\bm \theta}))$ for Model~2.
Unless otherwise stated,
all reported binary solutions, cost values,
and correlator distributions
are evaluated at ${\bm \theta}^*$,
whereas the relaxed loss values shown in the optimization traces
correspond directly to the continuously optimized objective.
This distinction is particularly important in the present setting,
where the relaxed loss can continue to improve
even when the decoded binary state remains unchanged.

Because the decoded binary solution
is not in general locally optimal,
we also examine a post-processing step
applied after decoding.
We tested both the local search procedure
adopted by Sciorilli \emph{et al.}~\cite{Sciorilli2025PCE}
and a greedy method,
and found that the greedy approach
consistently yielded better performance.
Accordingly, greedy post-processing
is adopted when post-processed results are reported.
Throughout this paper,
the term ``complete PCE workflow'' refers to
PCE-based initialization,
sign-based decoding,
and the subsequent greedy post-processing step.
Unless otherwise stated,
all reported PCE results correspond to this complete workflow.

In the greedy method,
a single-pass update is performed:
binary variables are ranked by their individual cost-reduction contributions,
and are sequentially flipped in this order,
with each variable updated at most once.

\section{Results}
\label{sec:numerical_results}

\begin{figure}[htb]
  \includegraphics[width=8.5cm]{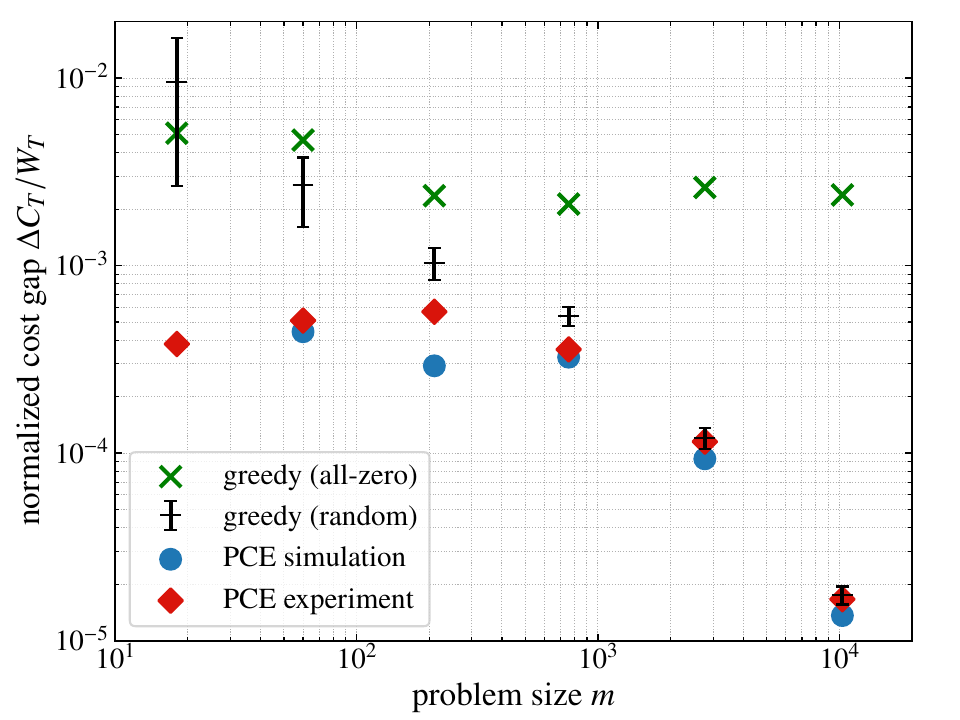}
    \caption{\label{fig:cost_convergence}
      Normalized cost gap as a function of the problem size $m$ for the time-averaged portfolio optimization (Model~1).
      The vertical axis shows the normalized cost gap $\Delta C_T / W_T$ [Eq.~(\ref{eq:normal_cost})].
      Crosses [greedy (all-zero)] denote results obtained by the greedy algorithm initialized from the all-zero bit string,
      while plus markers [greedy (random)] show results averaged over 1{,}000 runs with independent random initial bit strings;
      error bars indicate standard deviations over these runs.
      Circles and diamonds correspond to results obtained using the complete PCE workflow, including greedy post-processing after sign-based decoding,
      in simulation and on the quantum processing unit (QPU), respectively.
      For $m = 18$, the PCE solver achieves the exact optimal cost,
      yielding a zero normalized cost gap; the corresponding data point is not shown.
      The QPU results are obtained using circuit parameters optimized via variational quantum circuit simulation.
}
\end{figure}

\begin{table}[t]
  \centering
  \caption{
Summary of optimization settings and convergence statistics for Model~1.
Here, $m$ denotes the problem size (number of consumers),
$n$ the number of qubits, and $k$ the order of Pauli correlation operators.
$n_{\mathrm{params}}$ is the number of variational parameters.
The hyperparameters $\alpha_{\mathrm{sc}}$, $\beta$, and $\nu$
appear in the loss function (see text),
where $\nu$ corresponds to the Frobenius norm
of the cost matrix and has units of MWh$^2$.
The parameter $\alpha$ is defined as
$\alpha = \alpha_{\mathrm{sc}} n^{\lfloor k/2 \rfloor}$
following Ref.~\cite{Sciorilli2025PCE}.
The values of $\alpha_{\mathrm{sc}}$ and $\beta$
are selected to minimize the decoded cost $C_T$,
while $\nu$ is fixed.
For each instance, the optimization is repeated
over five independent runs
with different random initial circuit parameters.
The reported values correspond to the run
that yields the minimum decoded cost $C_T$,
and $n_{\mathrm{iter}}$ denotes
the number of optimization iterations for that run.
  }
    \label{tab:optimization_results}
    \begin{ruledtabular}
      \begin{tabular}{rrrrrrrr}
        $m$ & $n$ & $k$ & $\alpha_{\mathrm{sc}}$ & $\beta$ & $\nu$ &
$n_{\mathrm{params}}$ & $n_{\mathrm{iter}}$ \\
	\hline
          18 &  4 & 2 & 1.5 & 0.1 & $0.00421$ &  70 & 173 \\
          60 &  6 & 3 & 0.1 & 0.1 & $0.0320$ & 135 & 458 \\
         210 &  8 & 4 & 0.1 & 0.1 & $0.120$ & 220 & $1{,}291$ \\
         756 & 10 & 5 & 0.5 & 0.0 & $0.901$ & 325 & $1{,}450$ \\
        $2{,}772$ & 12 & 6 & 0.1 & 0.0 &  $6.52$ & 450 & $2{,}486$ \\
       $10{,}296$ & 14 & 7 & 0.1 & 0.0 &  $45.6$ & 595 & $2{,}609$ \\
    \end{tabular}
    \end{ruledtabular}
\end{table}

We first evaluate the time-averaged optimization problem (Model~1),
which serves as the initial stage of the hierarchical optimization framework.
Solving Model~1 provides variational circuit parameters
that capture the dominant structure of the cost landscape
and are used to initialize the subsequent time-dependent optimization (Model~2).

\subsection{Qubit Efficiency and Optimization Performance of PCE (Model~1)}
\label{sec:scalability}

Figure~\ref{fig:cost_convergence} shows the normalized cost gap for Model~1
as a function of the problem size $m$,
demonstrating the optimization performance of the PCE-based framework across a wide range of problem sizes.
The corresponding optimization settings and convergence statistics
are summarized in Table~\ref{tab:optimization_results}.

Across all tested instances,
the PCE-based solver achieves consistently low normalized cost gaps,
indicating that the proposed encoding and optimization strategy
remain effective as the problem size increases.
This scalability is enabled by the compact representation
of variables through Pauli correlation operators,
which avoids the linear growth of qubit requirements
typical of direct binary encodings.

The hyperparameters $\alpha_{\mathrm{sc}}$ and $\beta$
are selected to minimize the total cost $C_T$
via a grid search.
Here, $\alpha_{\mathrm{sc}}$ controls the scaling
$\alpha = \alpha_{\mathrm{sc}} n^{\lfloor k/2 \rfloor}$~\cite{Sciorilli2025PCE}.
Their dependence on system size is detailed in
Appendix~\ref{sec:alpha_beta}.
In particular, $\beta>0$ is beneficial for smaller instances,
whereas $\beta=0$ yields more stable performance for larger systems.

To assess the role of post-processing,
we compare against a greedy baseline initialized from random bit strings
[greedy (random)],
using the same single-pass update rule
as in Sec.~\ref{subsubsec:decoding_postprocessing}.
Since PCE also applies this greedy step after decoding,
this comparison isolates the quality of the binary initialization.
The corresponding cost distributions are shown in
Appendix~\ref{sec:cost_hist_appendix}.

The size dependence in Fig.~\ref{fig:cost_convergence}
can be understood from the intrinsic structure of the problem.
Here, the statistical quantities
$\mathbb{E}[p_{t,i}]$ and $\mathrm{Cov}(p_{t,i},p_{t,j})$
in Eq.~(\ref{eq:hourly_cost}) are fixed,
so the task reduces to a combinatorial optimization problem.
As $m$ increases, the aggregated quantity
$\sum_i p_{t,i} x_i$
may exhibit averaging effects,
which can reduce the influence of individual consumer fluctuations
on the overall portfolio behavior.
At the same time, the number of near-optimal configurations may increase,
leading to a higher degeneracy of solutions
with similar cost values.
These effects can make the optimization landscape
effectively smoother,
allowing high-quality solutions to be obtained more easily.
This interpretation is consistent with the baseline behavior:
while greedy (all-zero) shows a size-independent lower bound,
greedy (random) improves with increasing $m$.
This indicates that the performance gain in larger systems
is not solely algorithmic,
but reflects the increasing smoothness and redundancy
of the effective solution space.
This picture is further supported by the match probability analysis
in Sec.~\ref{sec:hardware_validation},
which shows that low-cost solutions increasingly differ
at the level of individual binary variables,
indicating an increase in the number of distinct near-optimal configurations.

Across all problem sizes,
the PCE simulation results remain consistently below
the greedy (random) baseline,
demonstrating that PCE provides better initial configurations.
For QPU experiments,
the same trend is observed up to $m=756$,
indicating that useful problem-dependent structure
is preserved in hardware-generated states.
For larger instances ($m = 2{,}772$ and $10{,}296$),
the QPU results approach the mean of the greedy (random) distribution,
suggesting that the initialization becomes effectively random
at the level of post-processed cost.
This behavior is consistent with the correlator distributions in
Appendix~\ref{sec:correlator_appendix}
and with the match probability analysis in
Sec.~\ref{sec:hardware_validation},
where the overlap between simulation and QPU solutions
approaches the random baseline.
These QPU results are obtained using parameters
optimized in noiseless simulations,
yet they follow the same overall trends,
indicating that the proposed method remains viable
on current quantum hardware.
Further details are provided in
Sec.~\ref{sec:hardware_validation}
and Appendix~\ref{sec:qpu_appendix}.

\subsection{Optimization Mechanism}
\label{sec:optimization_mechanism}

\begin{figure}[htb]
  \includegraphics[width=8.5cm]{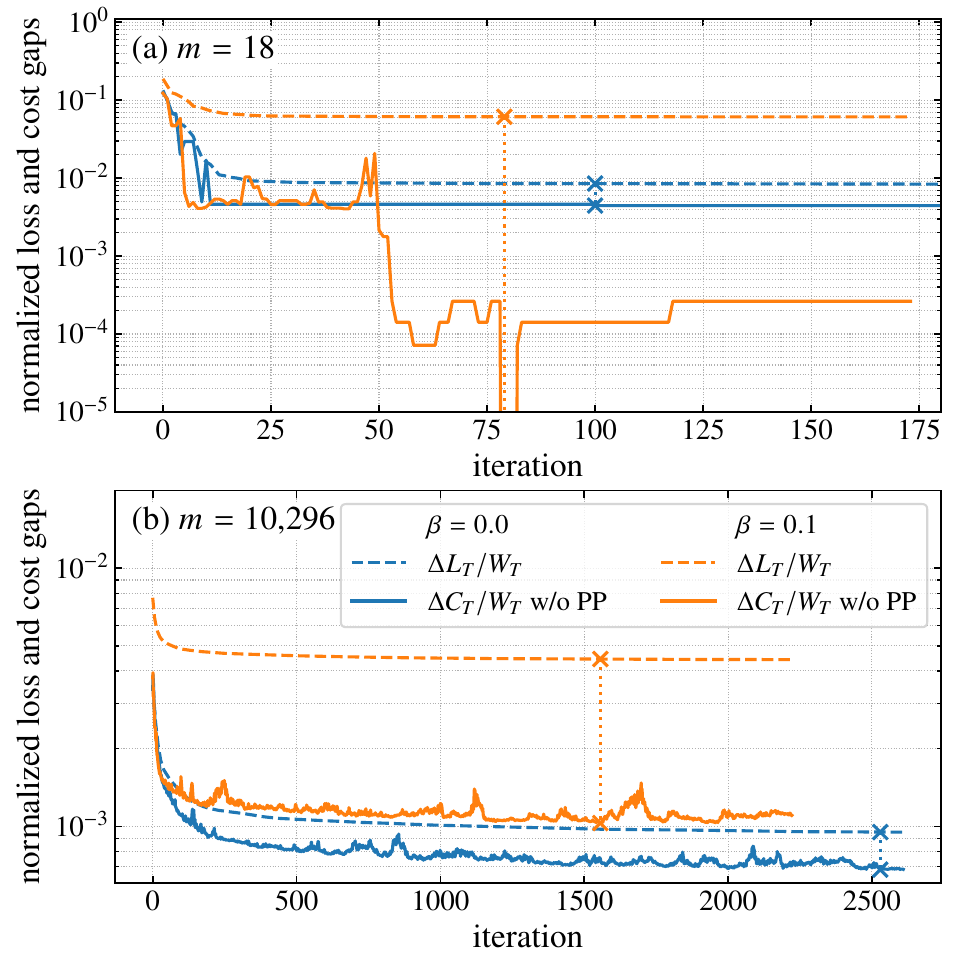}
\caption{\label{fig:ite_process}
Iteration dependence of the normalized loss $\Delta L_T/W_T$
and cost $\Delta C_T/W_T$ gaps (without post-processing, w/o PP).
Results are shown for $\beta=0.0$ and $0.1$.
Panels correspond to
(a) $m = 18$ with $\alpha_{\rm sc}=1.5$ ($\alpha=6.0$), and
(b) $m = 10{,}296$ with $\alpha_{\rm sc}=0.1$ ($\alpha=274.4$).
The continuous loss decreases monotonically during optimization,
whereas the corresponding decoded binary cost
exhibits step-like or irregular, non-monotonic behavior
due to the nonlinear discretization into binary variables.
Crosses connected by dotted lines indicate
the minimum cost values attained along the optimization trajectory,
together with the corresponding loss values
at the same iterations.
For $m=18$, the cost reaches zero around iteration $\sim 80$
and is partially truncated due to the logarithmic scale.
All results are obtained from state-vector simulations.
}
\end{figure}

To understand the relationship between the continuous optimization
and the discretized binary objective,
we analyze the optimization behavior
for representative small and large instances,
as shown in Fig.~\ref{fig:ite_process}.

For both system sizes,
the normalized loss gap decreases smoothly during optimization,
reflecting the continuous nature of the variational objective.
In contrast, the corresponding decoded binary cost
exhibits step-like or irregular, non-monotonic behavior
due to the nonlinear mapping
from continuous variables to binary variables.

\begin{figure}[htb]
  \includegraphics[width=8.5cm]{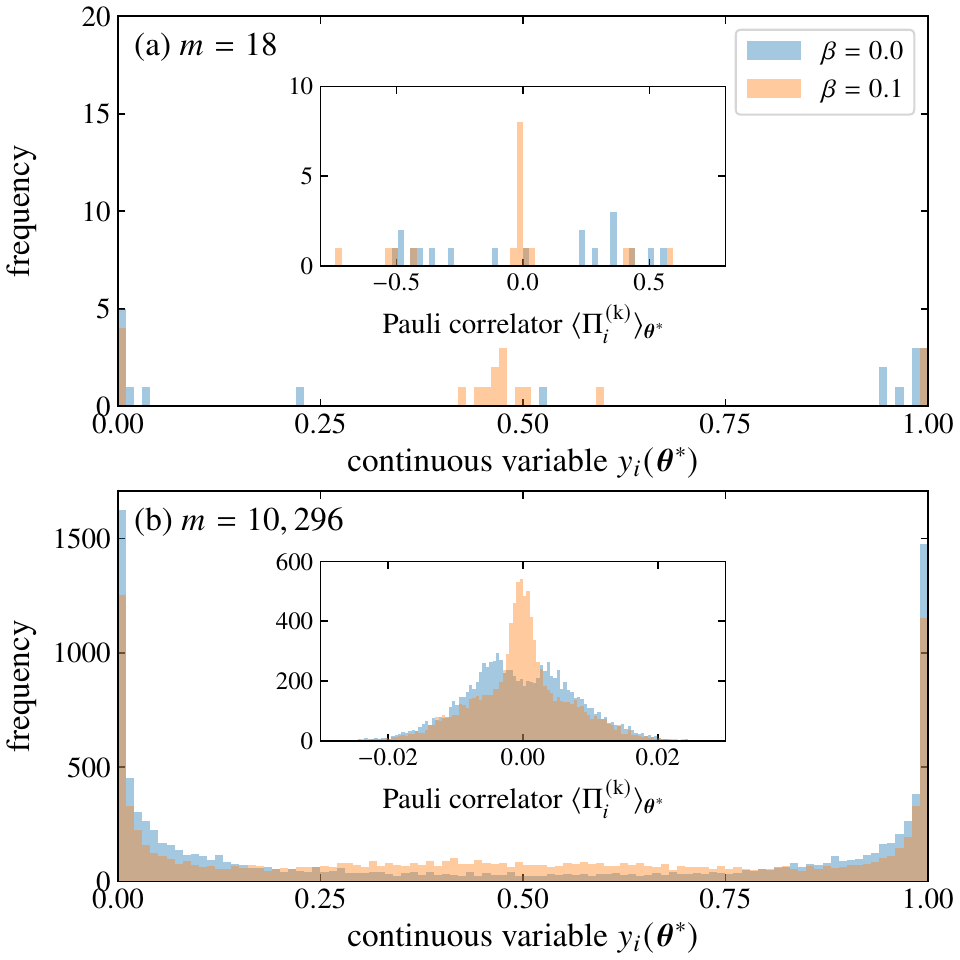}
    \caption{\label{fig:correlator_yi}
Frequency distributions of the sets $\{y_i({\bm \theta}^*)\}_{i=1}^m$ with
$y_i({\bm \theta}^*) = \varsigma(2\alpha_{\rm sc} \langle \Pi_i^{\rm (k)} \rangle_{{\bm \theta}^*})$.
Results are shown for $\beta=0.0$ and $0.1$.
Panels correspond to
(a) $m = 18$ with $\alpha_{\rm sc}=1.5$ ($\alpha=6.0$), and
(b) $m = 10{,}296$ with $\alpha_{\rm sc}=0.1$ ($\alpha=274.4$).
The main panels show the distributions of $\{y_i({\bm \theta}^*)\}_{i=1}^m$,
while the insets display those of the corresponding sets
$\{\langle \Pi_i^{\rm (k)} \rangle_{{\bm \theta}^*}\}_{i=1}^m$.
All distributions are evaluated at ${\bm \theta}^*$,
which corresponds to the minimum decoded binary cost
in Fig.~\ref{fig:ite_process} (without post-processing).
All results are obtained from state-vector simulations.
}
\end{figure}

To clarify the origin of this behavior,
we examine the distributions of the continuous variables
and the corresponding Pauli correlators,
shown in Fig.~\ref{fig:correlator_yi}.
These distributions are evaluated at the iterations
that achieve the minimum decoded binary cost values
in Fig.~\ref{fig:ite_process},
thereby characterizing the structure of the solutions
that are optimal after discretization
prior to post-processing.

For the small instance ($m=18$),
the optimization exhibits plateau-like cost behavior
together with effectively discrete correlator values,
as shown in Fig.~\ref{fig:ite_process}(a)
and Fig.~\ref{fig:correlator_yi}(a).
In this regime,
the regularization parameter $\beta$
strongly affects the optimization behavior.
Setting $\beta>0$ keeps the correlators near zero,
so that many variables remain close to the decoding boundary
after the nonlinear transformation.
As a result,
small fluctuations in the continuous variables
can trigger frequent changes in the discretized configuration,
leading to irregular cost variations across iterations,
which enhances exploration of the binary solution landscape.
In contrast, when $\beta=0$,
the correlators spread more broadly,
driving the continuous variables toward more polarized values.
As a result, the discretized configuration becomes stable but rigid,
and the decoded binary cost often stops improving
despite continued reduction of the relaxed loss.
This indicates that, in the small-system regime,
the bottleneck lies in transmitting continuous updates
into discrete binary variables.

For the large instance ($m=10{,}296$),
the optimization becomes substantially smoother after discretization,
and the correlator distributions become effectively continuous,
as shown in Fig.~\ref{fig:ite_process}(b)
and Fig.~\ref{fig:correlator_yi}(b).
In this regime,
the interplay between the continuous loss
and the decoded binary cost is reduced,
and the optimization behaves more consistently
with the relaxed objective.
Importantly, the role of $\beta$
changes systematically with system size.
For large systems,
transitions in the decoded binary variables
occur naturally due to the continuous correlator distribution,
and explicit regularization is no longer required.
Setting $\beta=0$ avoids biasing the variables
toward intermediate values,
thereby yielding more stable and lower-cost configurations.

Overall, these results show that
the optimization behavior in Model~1
is governed by the effective resolution
of the correlator representation,
which is ultimately controlled by the problem size $m$.
As the system size increases,
the representation transitions
from sparse and effectively discrete
to dense and effectively continuous.
Accordingly, the role of $\beta$
changes from promoting binary state transitions
in the small-$m$ regime
to becoming unnecessary, and eventually detrimental,
once continuous updates can be transmitted reliably
to decoded binary solutions.
This interpretation is supported by the hyperparameter dependence
shown in Appendix~\ref{sec:alpha_beta},
where run-to-run variability decreases with increasing $m$.

\subsection{Quantum Hardware Validation}
\label{sec:hardware_validation}

\begin{table}[t]
\centering
\caption{
  Comparison of normalized cost and loss gaps for the time-averaged portfolio optimization (Model~1) across different problem sizes $m$.
The columns show the normalized cost gap
$\Delta C_T/W_T$ with post-processing (w/ PP),
the normalized cost gap $\Delta C_T/W_T$
without post-processing (w/o PP),
and the normalized loss gap $\Delta L_T/W_T$.
Results are shown for both quantum circuit simulation and QPU implementations.
Post-processing improves the normalized cost gap, particularly for QPU results, while the normalized loss gap remains similar between simulation and QPU.
  }
\label{tab:hardware_feasibility}
\begin{ruledtabular}
\begin{tabular}{r c c c c}
$m$
  &\makecell{evaluation\\ setting}
& \makecell{$\Delta C_T/W_T$ \\ w/ PP}
& \makecell{$\Delta C_T/W_T$ \\ w/o PP}
& $\Delta L_T/W_T$   \\
\hline
\multirow{2}{*}{$18$}
& simulation 
& $0.0$
& $0.0$
& $6.13 \times 10^{-2}$ \\
& experiment
& $3.82 \times 10^{-4}$
& $2.93 \times 10^{-2}$
& $6.78 \times 10^{-2}$ \\
\\
\multirow{2}{*}{$60$}
& simulation 
& $4.44 \times 10^{-4}$
& $6.47 \times 10^{-4}$
& $3.46 \times 10^{-2}$ \\ 
& experiment
& $5.10 \times 10^{-4}$
& $4.00 \times 10^{-3}$
& $3.61 \times 10^{-2}$ \\
\\
\multirow{2}{*}{$210$}
& simulation
& $2.92 \times 10^{-4}$
& $5.44 \times 10^{-4}$
& $9.96 \times 10^{-3}$ \\ 
& experiment
& $5.68 \times 10^{-4}$
& $2.97 \times 10^{-3}$
& $1.36 \times 10^{-2}$ \\
\\
\multirow{2}{*}{$756$}
& simulation
& $3.25 \times 10^{-4}$
& $4.83 \times 10^{-4}$
& $6.78 \times 10^{-4}$ \\ 
& experiment
& $3.58 \times 10^{-4}$
& $2.72 \times 10^{-3}$
& $4.19 \times 10^{-3}$ \\
\\
\multirow{2}{*}{$2{,}772$}
& simulation
& $9.34 \times10^{-5}$
& $3.53 \times10^{-4}$ 
& $5.33 \times10^{-4}$ \\ 
& experiment
& $1.15 \times10^{-4}$
& $3.90 \times10^{-3}$
& $3.91 \times10^{-3}$ \\
\\
\multirow{2}{*}{$10{,}296$}
& simulation
& $1.36 \times10^{-5}$
& $6.80 \times10^{-4}$ 
& $9.51 \times10^{-4}$ \\
& experiment
& $1.67 \times10^{-5}$
& $3.04 \times10^{-3}$
& $2.98 \times10^{-3}$ \\
\end{tabular}
\end{ruledtabular}
\end{table}

\begin{table}[t]
\centering
\caption{
  Match probability (fraction of identical components)
  between portfolios obtained by
  quantum circuit simulation (PCE simulation),
  QPU experiments (PCE experiment),
  and optimal solutions (optimal) for Model~1,
  evaluated across different problem sizes $m$.
  For the PCE-based method,
  the portfolios correspond to the complete PCE workflow,
  including sign-based decoding and greedy post-processing.
  The optimal solution was obtained using the commercial solver Gurobi (version 12.0.3)~\cite{gurobi},
  with the MIP optimality tolerance (relative gap $10^{-4}$).
}
\label{tab:match_probability}
\begin{ruledtabular}
\begin{tabular}{rccc}
size $m$
& \makecell{PCE simulation \\ vs. \\ optimal}
& \makecell{PCE simulation \\ vs. \\ PCE experiment}
& \makecell{PCE experiment \\ vs. \\ optimal}\\
\hline
18    & 1.0000 & 0.6667 & 0.6667 \\
60    & 0.7000 & 0.7833 & 0.6167 \\
210   & 0.6905 & 0.6238 & 0.5714 \\
756   & 0.7606 & 0.6852 & 0.6071 \\
2{,}772  & 0.7691 & 0.5278 & 0.5191 \\
10{,}296 & 0.6527 & 0.5125 & 0.5125 \\
\end{tabular}
\end{ruledtabular}
\end{table}

In addition to the simulation results,
we perform experiments on a quantum processing unit (QPU)
to assess the practical feasibility of the proposed framework.
All experiments are carried out using circuit parameters ${\bm \theta}^*$
optimized in noiseless state-vector simulations,
thereby isolating the effect of hardware noise
from that of variational parameter optimization.

The experiments are conducted on the IonQ Forte QPU
available via Amazon Braket~\cite{AmazonBraket}.
This device is based on trapped-ion technology
and provides all-to-all qubit connectivity,
which is well suited to the present implementation,
as the Pauli correlation encoding (PCE)
involves nonlocal interactions among qubits.

For each problem instance,
the optimized circuit parameters are fixed,
and a single round of expectation-value measurement,
decoding, and cost evaluation is performed,
without any feedback or parameter updates on hardware.
This setup enables a direct comparison
between ideal simulation and hardware execution.
To ensure consistent statistical precision,
a fixed number of 4{,}001 shots is used
for each measurement basis.
The Pauli correlators are grouped into mutually commuting sets
corresponding to the $X$, $Y$, and $Z$ bases,
and measurements are performed independently for each basis,
resulting in a total of $3 \times 4{,}001$ shots per circuit execution.

The resulting performance is summarized in Table~\ref{tab:hardware_feasibility},
which reports the normalized cost gap $\Delta C_T/W_T$
with post-processing (w/ PP),
the normalized cost gap $\Delta C_T/W_T$
without post-processing (w/o PP),
and the normalized loss gap $\Delta L_T/W_T$
for both simulations and QPU experiments.
We observe that the normalized loss gap remains close
between simulation and QPU across all problem sizes,
indicating that the expectation values of the parametrized circuits
are largely preserved under hardware noise.
In contrast, the cost without post-processing
is significantly degraded in the QPU results,
indicating that the decoding step is sensitive
to sampling noise and small fluctuations
in the continuous variables.
This sensitivity is particularly pronounced
for correlators located near zero,
because sign-based decoding
can change discontinuously
under small estimation errors.
Specifically, in the present experiments,
each measurement basis was sampled using
$N_{\mathrm{shot}} = 4{,}001$ shots.
Under finite-shot sampling,
the statistical uncertainty of an estimated correlator
scales as $O(1/\sqrt{N_{\mathrm{shot}}})$,
corresponding to a characteristic resolution of
approximately $1/\sqrt{4001}\approx 0.0158$.
Consequently, correlators whose expectation values satisfy
$|\langle\Pi_i^{(k)}\rangle_{\bm \theta}|
\lesssim 0.016$
are particularly sensitive to statistical fluctuations,
which can induce sign errors during decoding.
This observation provides a quantitative explanation
for the increased sensitivity of the decoded QPU solutions
to finite-shot noise.
It is also consistent with the correlator distributions
shown in Appendix~\ref{sec:correlator_appendix},
where noticeable deviations from the ideal state-vector
distributions are observed in the QPU results,
particularly for large problem sizes.
Such deviations are consistent with the combined effects
of finite-shot fluctuations and hardware imperfections,
which can perturb correlator values near the decoding boundary
and increase the probability of sign errors during decoding.
Consequently,
correlators close to the decoding boundary
require a larger number of measurement shots
to determine their signs reliably.

Importantly, applying greedy post-processing
substantially improves the quality of the decoded solutions,
bringing the normalized cost gap close to the corresponding simulation results
for all problem sizes.
This demonstrates that the combination of PCE
and post-processing
can effectively mitigate the impact of hardware noise
on the final binary solutions.

\begin{figure*}[htb]
  \includegraphics[width=15cm]{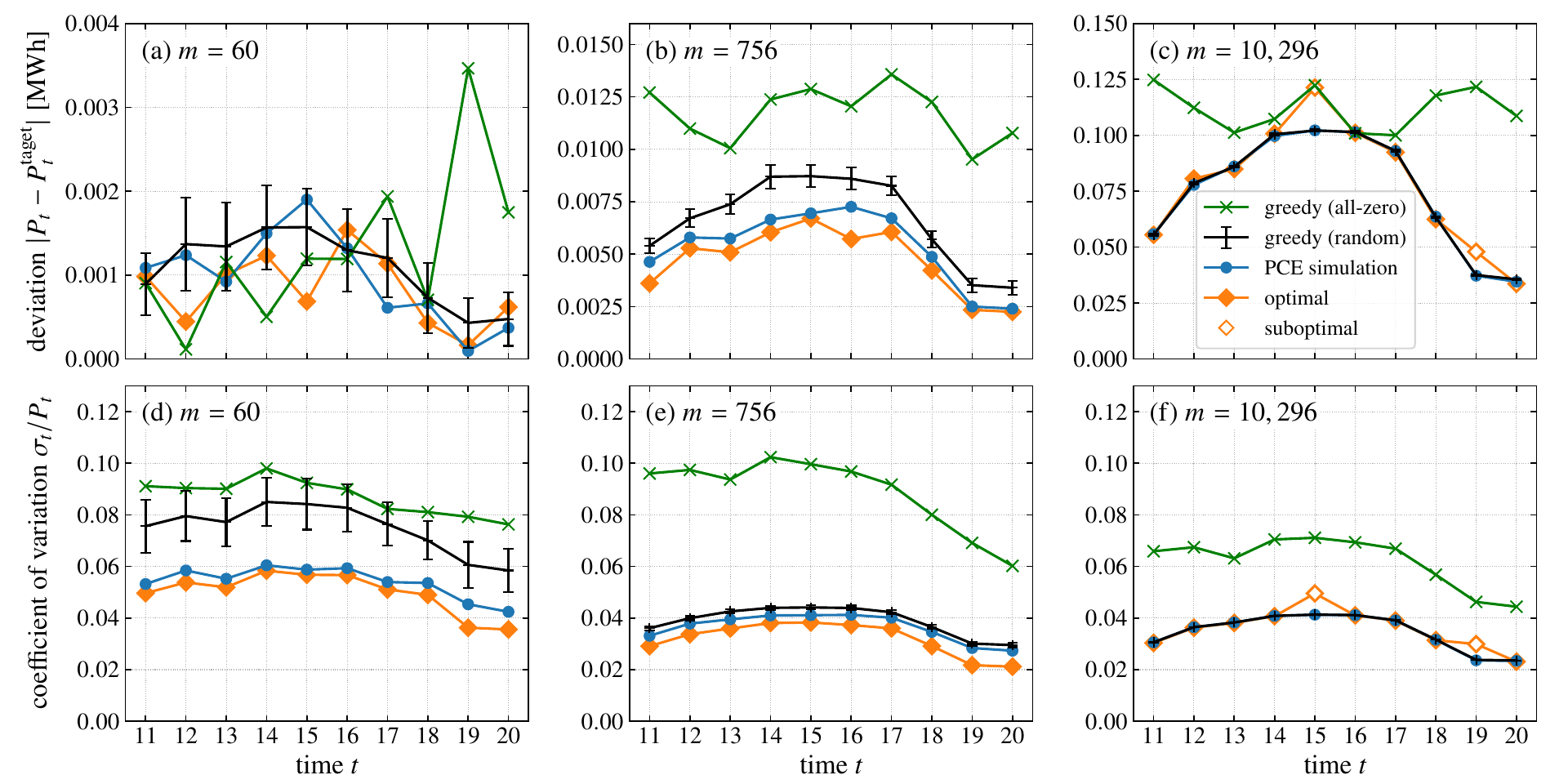}
    \caption{\label{fig:deviation_and_CV}
      Time-resolved deviation from the target power and coefficient of variation for three problem sizes ($m = 60$, $m = 756$, and $m = 10{,}296$).
    (a),(b),(c) $|P_t - P^{\mathrm{target}}_t|$.
    (d),(e),(f) $\sigma_t / P_t$, where $\sigma_t$ is the standard deviation at time $t$.
      Panels (a),(d)  correspond to $m = 60$, (b),(e) to $m = 756$, and (c),(f) to $m=10{,}296$.
          Results obtained by the greedy algorithm include those from the all-zero initialization [greedy (all-zero)] and
    from random initialization [greedy (random)] averaged over 1{,}000 runs with independent random initial bit strings;
    error bars indicate standard deviations over these random-initialization runs.
The optimal solutions were computed using the commercial solver Gurobi (version 12.0.3)~\cite{gurobi}
with the MIP optimality tolerance (relative gap $10^{-4}$).
For $m=10{,}296$, the solver did not reach convergence within a 24-hour time limit
at $t=15, 19$, and $20$.
In these cases, the suboptimal solutions are shown with white diamond markers.
The optimality gaps remain large for $t=15$ and $19$, while the gap at $t=20$ is below $0.1\%$.
For the PCE-based method,
all results are obtained from the complete PCE workflow,
including greedy post-processing after sign-based decoding.
    }
\end{figure*}

To further assess the consistency of the obtained solutions,
we evaluate the match probability between portfolios
obtained from simulations, QPU experiments, and optimal solutions,
as summarized in Table~\ref{tab:match_probability}.
Each portfolio is represented as a binary vector
${\bm x} \in \{0,1\}^m$,
and the match probability is defined as the fraction of components
that take identical values between two vectors.

The match probability between simulation and QPU results
is moderate for small to intermediate problem sizes,
indicating that a non-negligible fraction of binary variables
is preserved under hardware noise,
although discrepancies in individual variables persist.
For larger instances ($m = 2{,}772$ and $10{,}296$),
the match probability decreases to values close to 0.5,
which corresponds to the expected agreement between
approximately balanced, uncorrelated binary configurations
in the present problem setting.
This indicates that the correspondence between configurations
becomes nearly indistinguishable from random agreement.

Furthermore, the match probability between simulation and optimal solutions
is consistently higher than that between QPU experiments and optimal solutions,
suggesting that the degradation observed in hardware primarily arises
from sampling noise and measurement errors,
rather than from the variational ansatz itself.
In particular, the match probability between simulation and optimal solutions
remains in the range of approximately $0.65$--$0.77$ for larger problem sizes,
indicating that the variational optimization successfully captures
a large fraction of the optimal binary structure.

Importantly, despite the relatively low match probability,
the corresponding normalized cost gaps remain close to optimal values,
especially for larger problem sizes.
This indicates that multiple distinct binary configurations
can yield similarly low cost,
which is consistent with the increasing degeneracy of near-optimal solutions
discussed in Sec.~\ref{sec:scalability}.

Taken together, these observations indicate that,
while hardware noise affects the fidelity of individual binary variables,
the overall structure of the optimized portfolios
is reasonably well preserved,
and the proposed framework remains effective
in producing high-quality approximate solutions on current quantum hardware.

\subsection{Time-Resolved Portfolio Accuracy and Stability (Model~2)}

We next evaluate the time-dependent optimization (Model~2).
For each instance, we use the same parameter settings as in Table~\ref{tab:optimization_results} for Model~1,
except for the number of iterations $n_{\mathrm{iter}}$.
The circuit parameters ${\bm \theta}^*$ obtained in Model~1 serve as an initialization,
but are further updated via variational optimization
with respect to the time-resolved objective in Eq.~(\ref{eq:hourly_cost}).
The resulting solutions are then converted to binary form
and refined by greedy post-processing.

Figure~\ref{fig:deviation_and_CV} summarizes the deviation from the procurement target
and stability through $|P_t - P^{\mathrm{target}}_t|$ and $\sigma_t / P_t$
for $m = 60$, $756$, and $10{,}296$.
For $m = 60$ and $756$,
the PCE-based method closely follows the optimal solution
in time slots where optimality is verified,
while consistently outperforming the greedy (random) baseline
in both deviation and variability.

The relative improvement depends on the problem size.
For $m = 60$, differences in deviation are small,
but the coefficient of variation (CV) reveals a clear reduction in variability.
For $m = 756$, a clear reduction in deviation is observed,
whereas the CV remains comparable between the PCE-based method
and the greedy (random) baseline,
indicating that the primary improvement appears in accuracy
rather than variability.

For the largest instance ($m = 10{,}296$),
the PCE-based solutions are nearly indistinguishable from both the optimal solutions
and the greedy (random) baseline in time slots where optimality is verified,
with differences too small to be visually resolved in the plots.
In contrast, the classical solver (Gurobi) does not reach certified optimality for time slots $t = 15$, $19$, and $20$.
Among these, the suboptimal solutions at $t = 15$ and $19$ exhibit significantly larger deviations and CV,
leading to clearly degraded performance.
The near-indistinguishability observed in the optimal time slots
indicates that, in the large-system regime,
the cost landscape becomes increasingly flat around the optimum,
so that many distinct configurations yield nearly identical cost values.
This behavior is consistent with the increasing degeneracy of near-optimal solutions
observed in Model~1.
As a result, solutions obtained from different methods
tend to converge to effectively equivalent performance.
In this regime, the benefit of structured initialization is reduced,
and even simple greedy refinement from random initial configurations
can approach high-quality solutions.

Overall, PCE provides reliable time-dependent solutions across system sizes,
with clear advantages in intermediate regimes.
These results highlight the role of variational optimization
in constructing structured continuous representations
that improve solution quality prior to binary decoding.

\section{Discussion}
\label{sec:discussion}

The present results clarify three main aspects
of Pauli correlation encoding (PCE)
for large-scale binary optimization.

First, PCE provides a qubit-efficient variational representation
for fully connected quadratic optimization problems
with real-valued and nonuniform coefficients.
In the present application to electric power demand portfolio optimization,
a large number of binary variables
is represented through expectation values
of Pauli correlation operators
acting on a comparatively small number of qubits.
This compression enables dense large-scale instances
to be treated without requiring a direct one-qubit-per-variable encoding.

Second, the optimization behavior of PCE
is fundamentally governed by the interplay
between the continuously relaxed loss landscape
and the discretized binary objective.
Optimization is performed over continuous variables
derived from Pauli correlators,
whereas the final solution is evaluated
after threshold-based decoding.
As a result, improvements in the relaxed loss
do not always translate into improvements
in the decoded binary cost.
To clarify the underlying mechanism of this interplay,
it is instructive to analyze its system-size dependence.
For small systems, the correlator representation is effectively low-resolution,
and continuous updates do not always induce changes
in the decoded binary state.
For large systems, the distributions become effectively continuous,
allowing updates in the relaxed variables
to be transmitted more reliably to the binary solution.
Consequently, the optimization becomes increasingly consistent
with the binary objective.
This mechanism also explains the size-dependent role of regularization.
While finite regularization can facilitate transitions
between binary configurations in small systems,
it becomes unnecessary or even detrimental in larger systems,
where the representation already provides sufficient effective resolution.
The decreasing run-to-run variability with increasing system size
further supports this interpretation.
Importantly, this resolution-dependent behavior
also explains the scalability observed in
Sec.~\ref{sec:numerical_results}.
As the system size increases,
the optimization landscape may become effectively smoother,
and the number of near-optimal configurations may increase.
This is consistent with both the decrease in normalized cost gap
and the match probability analysis,
which shows that distinct binary solutions
can yield similarly low cost values for large problem sizes.
We emphasize that
the observed improvement with increasing problem size
does not necessarily imply
that the optimization problem becomes intrinsically harder
and is subsequently solved more effectively.
Indeed,
averaging effects associated with large portfolios,
together with the increasing degeneracy
of near-optimal configurations,
may make large instances easier
from the viewpoint of optimization.
Therefore,
the improved performance observed for large problem sizes
should be interpreted together with these problem-dependent effects.
An additional open question concerns the relationship
between representational compression
and ansatz expressibility.
Although PCE enables a combinatorial increase
in the number of representable variables,
the extent to which the required expressibility
of the variational ansatz increases with
the correlation order $k$
or the number of encoded variables $m$
remains unclear.
In particular, maximizing the number of representable variables
through the choice $k=n/2$
may increase the complexity of the correlator space
that must be explored by a finite-depth variational circuit.
Understanding whether larger representational capacity
requires systematically deeper or more expressive ansatz
is therefore an important question.
A systematic investigation of the relationship among
correlation order,
ansatz depth,
expressibility,
and optimization performance
is beyond the scope of the present work
and remains an important direction for future research.

Third, the hardware results demonstrate that
the PCE representation retains useful problem-dependent structure
under finite sampling and device noise.
For small and intermediate systems,
QPU-generated bit strings outperform the greedy (random) baseline,
indicating that nontrivial information survives in the measured correlators.
For larger systems,
the correlator distributions approach random-like behavior,
and the decoded costs converge toward the greedy baseline.
This degradation is directly reflected in the correlator distributions,
which broaden and concentrate near zero under hardware noise.

Finally, the scaling advantage considered in this work
primarily concerns representational scalability,
namely the ability to represent a large number of optimization variables
using a comparatively small number of qubits.
This should be distinguished from
overall computational scalability,
which additionally depends on
measurement overhead,
objective-function evaluation cost,
and post-processing cost.
Accordingly, the present results should not be interpreted as
a demonstration of end-to-end computational scalability
or quantum advantage,
but rather as an investigation of the representational
and optimization properties of PCE.
For the present implementation,
all Pauli correlators can be estimated
from measurements in only three global bases
($X$, $Y$, and $Z$),
since the correlators are grouped into mutually commuting sets.
Consequently,
the number of distinct quantum state preparations and measurement configurations
remains strictly constant at three, regardless of the problem size $m$.
This property substantially reduces
the measurement-setting overhead
that typically scales with the number of terms
in conventional variational algorithms.
However,
the statistical precision required to determine correlator values
and their signs may increase as correlators approach zero,
potentially increasing the shot requirements
for reliable decoding in large-scale instances.
In addition,
the classical evaluation of the dense QUBO objective
scales as $O(m^2)$,
while the greedy post-processing procedure
scales as $O(m)$.
Therefore,
although PCE substantially reduces the number of required qubits,
the overall computational cost remains influenced by both
quantum measurement resources
and classical processing costs.

\section{Conclusion}
\label{sec:conclusion}

In this work, we have introduced
a representationally scalable variational framework
based on Pauli correlation encoding (PCE)
and demonstrated its effectiveness
for fully connected quadratic optimization problems
with real-valued and nonuniform coefficients,
using electric power demand portfolio optimization
as a representative application.
By representing binary variables
through expectation values of Pauli correlation operators,
the proposed approach enables a compact encoding
of a large number of variables
within a limited number of qubits,
while retaining sufficient information
to recover high-quality discrete solutions.

A central finding of this study
is that the optimization behavior of PCE
is governed by the interplay
between continuous relaxation
and discrete binary structure.
By analyzing the distribution of Pauli correlators,
we have shown that the effective resolution
of the correlator representation
determines how reliably improvements
in the continuous loss landscape
are reflected in the decoded binary solutions.
In particular, the transition
from sparse, effectively discrete distributions
to dense, effectively continuous ones
provides a unifying explanation
for the observed dependence
on system size and regularization.
This reveals that the success of the method
is not solely determined by optimization heuristics,
but is intrinsically linked
to the statistical structure of the underlying quantum state.

From a practical perspective,
we have demonstrated that the proposed framework
achieves near-optimal performance
over a wide range of problem sizes,
and remains robust under realistic conditions,
including finite sampling and hardware noise
on a trapped-ion quantum processor.
These results indicate that
the continuous relaxation induced by Pauli correlators
preserves sufficient information
about the discrete optimization landscape
to enable reliable solution recovery
even in non-ideal experimental settings.

More broadly, our results establish PCE
as a physically motivated and qubit-efficient framework
that connects combinatorial optimization
to the statistical structure of quantum states.
This perspective provides a principled way
to design variational algorithms
in which discrete variables are encoded
through measurable correlation functions,
offering a complementary approach
to conventional qubit-based encodings.
We expect that this framework
can be extended to a wider class of optimization problems
and may serve as a foundation
for developing qubit-efficient quantum optimization methods
in the NISQ regime and beyond.

\section*{Acknowledgments}
K.F. is supported by MEXT Quantum Leap Flagship Program (MEXTQLEAP) Grants No. JPMXS0118067394 and No. JPMXS0120319794.
This work is supported by JST COI-NEXT program Grant No. JPMJPF2014.

\appendix
\begin{figure*}[htb]
  \centering
  \includegraphics[width=15cm]{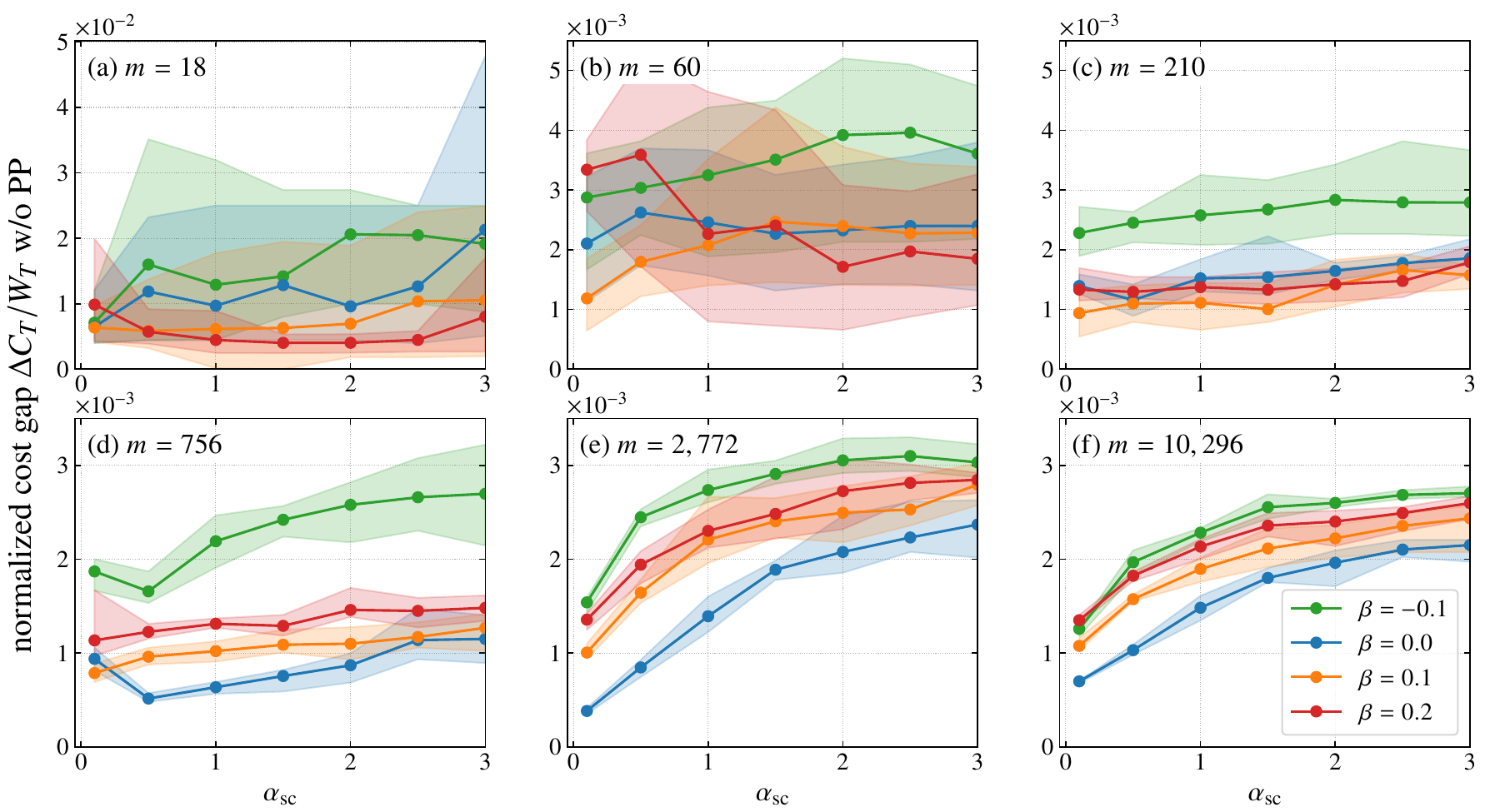}
  \caption{\label{fig:alpha_beta}
  Normalized cost gap $\Delta C_T / W_T$ without post-processing (w/o PP)
  for the time-averaged formulation (Model~1),
  shown as a function of the scaled parameter $\alpha_{\mathrm{sc}}$
  for different values of $\beta$ in the PCE solver.
  Panels (a)--(f) correspond to problem sizes
  $m=18, 60, 210, 756, 2{,}772,$ and $10{,}296$, respectively.
  For each parameter setting, the marker indicates
  the mean over five independent optimization runs
  with different random initial circuit parameters,
  while the shaded vertical range indicates
  the minimum and maximum values obtained across those runs.
  A clear size-dependent preference for $\beta$ is observed,
  with finite $\beta$ favored for small systems
  and $\beta=0$ for large systems.
  The parameter $\alpha$ controls the sharpness of the relaxation,
  while $\beta$ determines the strength of the quadratic regularization
  toward the continuously relaxed variables $y_i \in [0,1]$
  centered at $1/2$.
  }
\end{figure*}

\section{Hyperparameter Dependence of the Optimization Performance}
\label{sec:alpha_beta}

In this Appendix, we present a systematic exploration
of the hyperparameters $\alpha$ and $\beta$
in the PCE solver for Model~1.
The parameter $\alpha$ controls the sharpness
of the sigmoid relaxation
that maps Pauli correlators to continuous variables,
while $\beta$ determines the strength
of the quadratic regularization term
that biases the relaxed variables toward $y_i = 1/2$.

Figure~\ref{fig:alpha_beta} shows
the normalized cost gap $\Delta C_T / W_T$ without post-processing
as a function of the scaled parameter $\alpha_{\mathrm{sc}}$
for several values of $\beta$ and problem sizes.
The scaling
$\alpha_{\mathrm{sc}} = \alpha / n^{\lfloor k/2 \rfloor}$
is introduced to compare different system sizes
on an equal footing.

The results show that the optimization performance
depends sensitively on both $\alpha$ and $\beta$.
For small $\alpha$, the relaxation is too smooth
to induce effective binary separation,
whereas excessively large $\alpha$
can lead to saturation of the relaxed variables.
Similarly, too small $\beta$ provides insufficient regularization
in the small-system regime,
whereas too large $\beta$ can overconstrain the variables
around $y_i = 1/2$.

Importantly, the optimal choice of $\beta$
exhibits a clear size dependence.
For smaller instances ($m=18, 60,$ and $210$),
finite $\beta$ (e.g., $\beta=0.1$) consistently yields better performance,
whereas for larger instances
($m=756, 2{,}772,$ and $10{,}296$),
$\beta=0$ provides more stable and lower-cost solutions.
This trend is consistent with the mechanism discussed in the main text:
for small systems, sparse and effectively discrete correlator distributions
make the decoded binary state sensitive to small changes
in the relaxed variables,
whereas for large systems,
the correlator distributions become effectively continuous,
making explicit regularization unnecessary.
The same analysis also shows that the run-to-run variability
is substantially larger for small systems
and decreases with increasing problem size,
further supporting the interpretation
that the small-system regime is more strongly affected
by the interplay between the continuously relaxed loss
and the discretized binary objective.

A second notable feature is the strong size dependence
of the run-to-run variability.
For small systems,
the spread between the minimum and maximum values
across independent runs is substantial,
indicating strong sensitivity
to the initial circuit parameters.
By contrast, this spread becomes systematically smaller
as the problem size increases.
This behavior is consistent with the interpretation
that the optimization instability in small systems
originates not only from the loss landscape itself,
but from a mismatch between the continuously relaxed loss
and the discretized binary objective:
small changes in the relaxed variables
can lead to large differences after decoding.
As the system size increases and the correlator representation
becomes effectively denser,
this mismatch is progressively mitigated,
and the optimization becomes more reproducible
across independent runs.

Based on these observations,
we select the near-optimal values of $\alpha$ and $\beta$
summarized in Table~\ref{tab:optimization_results}
in the main text.

\section{Additional Analysis of QPU Results (Model~1)}
\label{sec:qpu_appendix}
In this Appendix, we provide additional quantitative analyses
of the results obtained from quantum circuit simulations
and QPU experiments.
These analyses complement the summary metrics presented in the main text
by explicitly examining the underlying distributions,
thereby clarifying the origin of the observed performance differences
between simulation and hardware implementations.

\begin{figure*}[htb]
  \centering
  \includegraphics[width=15cm]{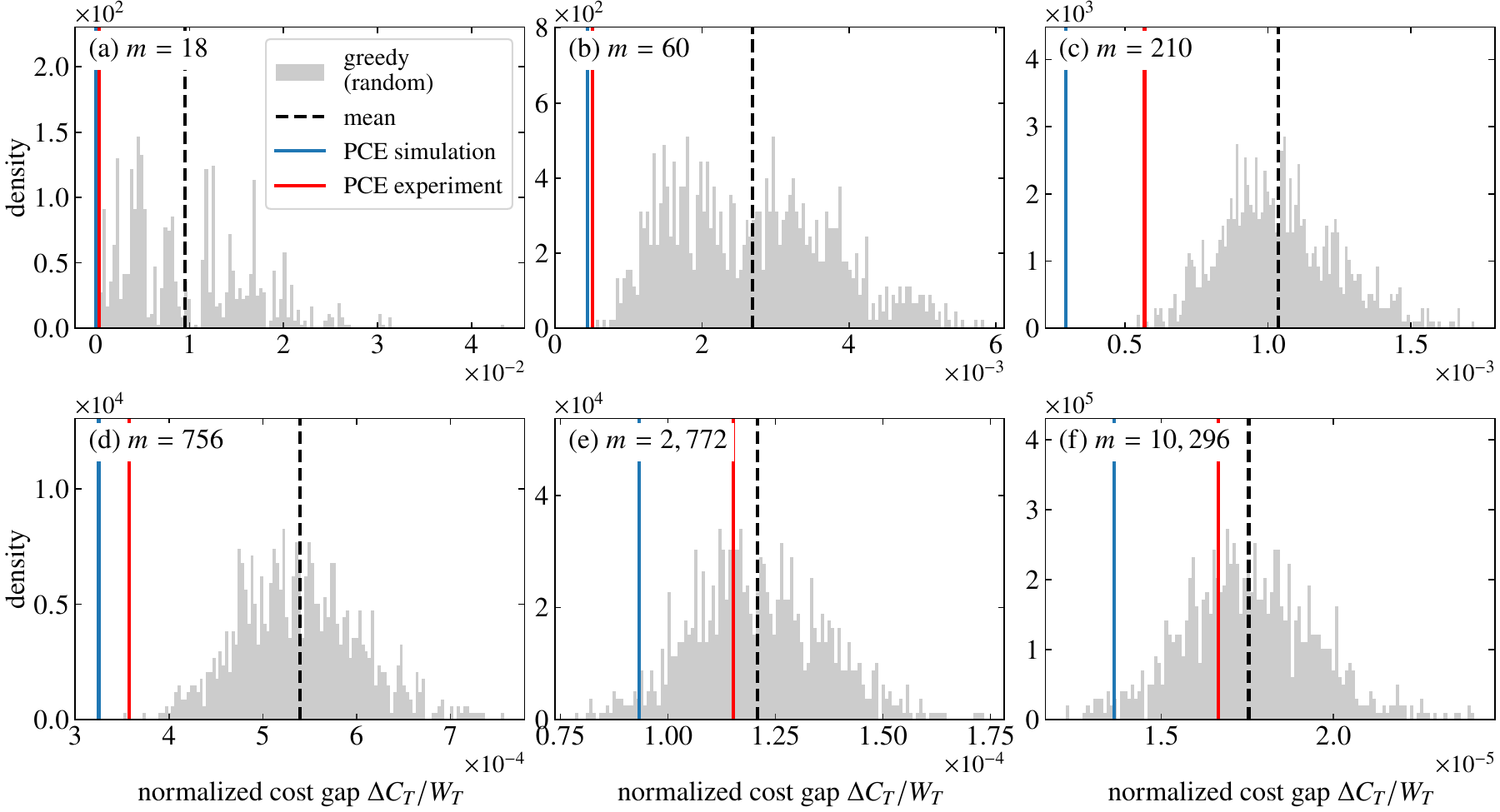}
  \caption{\label{fig:cost_histogram_6fig}
  Distributions of normalized cost gaps $\Delta C_T/W_T$
  for the time-averaged formulation (Model~1),
  for (a) $m = 18$, (b) $m = 60$, (c) $m = 210$,
  (d) $m = 756$, (e) $m = 2{,}772$, and (f) $m = 10{,}296$.
  The horizontal axis represents the normalized cost gap, and the vertical axis shows the frequency.
  The histograms correspond to the distributions obtained by applying a greedy algorithm
  to randomly initialized bit strings.
  The blue and red vertical lines indicate the normalized cost gaps
  $\Delta C_T/W_T$
  obtained from the complete PCE workflow,
  including greedy post-processing after sign-based decoding,
  for state-vector simulation and QPU experiment, respectively.
  }
\end{figure*}

\begin{figure*}[htb]
  \centering
  \includegraphics[width=15cm]{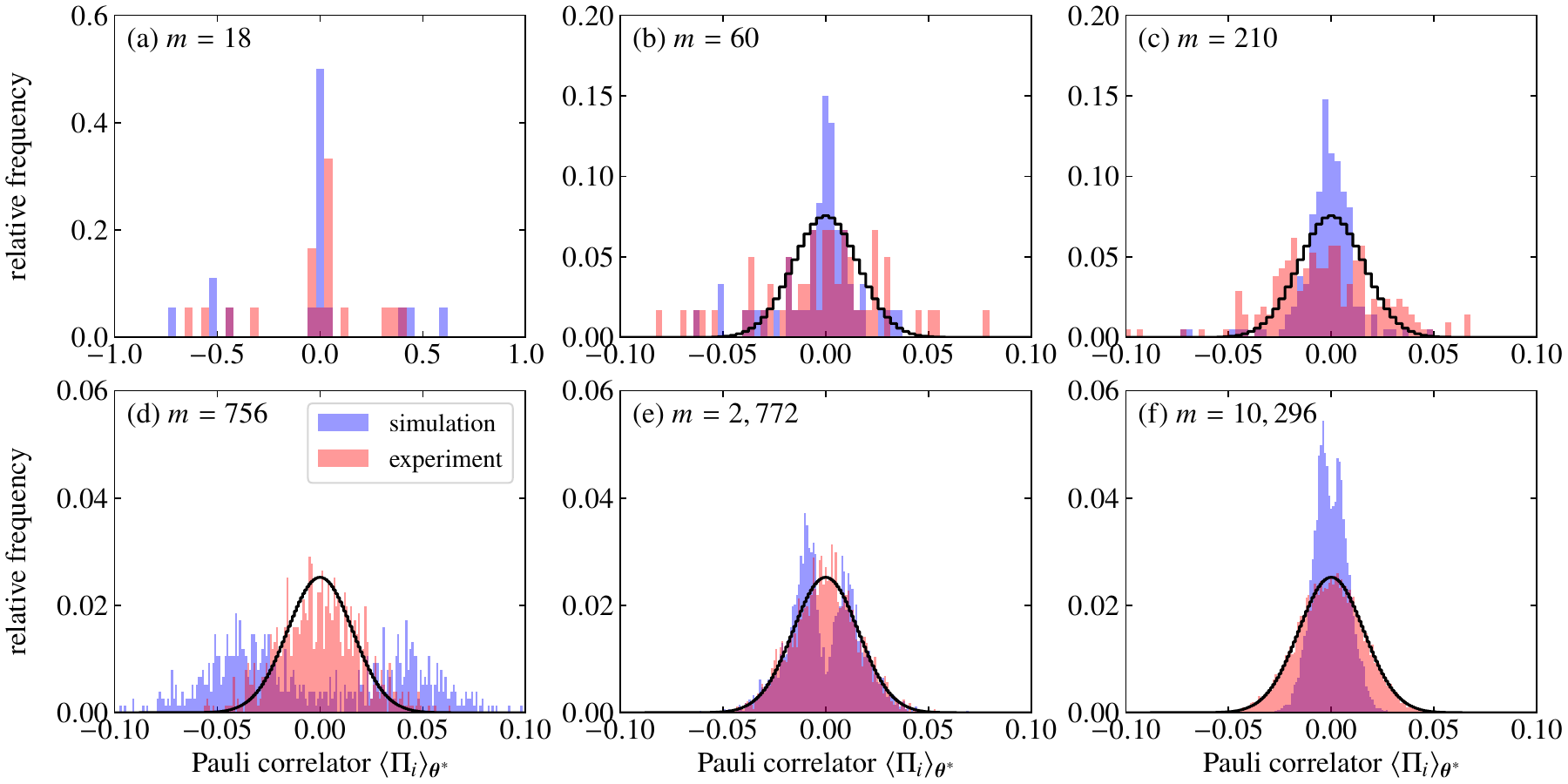}
  \caption{\label{fig:correlator_6fig}
  Histograms of the sets $\{\langle \Pi_i^{\rm (k)} \rangle_{{\bm \theta}^*}\}_{i=1}^m$
  for the time-averaged formulation (Model~1),
  for different problem sizes:
  (a) $m = 18$, (b) $m = 60$, (c) $m = 210$,
  (d) $m = 756$, (e) $m = 2{,}772$, and (f) $m = 10{,}296$.
  The parameters ${\bm \theta}^*$ are obtained using the hyperparameters
  $\alpha$ and $\beta$ listed in Table~\ref{tab:optimization_results}.
  All distributions are evaluated in the same manner as in Fig.~\ref{fig:correlator_yi}.
  Results obtained from state-vector simulations and QPU experiments are overlaid.
  For reference, the centered binomial distribution corresponding to independent
  $\{-1,1\}$ variables with probability $p = 0.5$ is also shown for sufficiently large instances,
  while it is omitted for the smallest case ($m = 18$) due to the limited number of correlators.
  }
\end{figure*}

\subsection{Cost Distributions}
\label{sec:cost_hist_appendix}

Figure~\ref{fig:cost_histogram_6fig} shows the distributions of normalized cost gaps
obtained by applying greedy post-processing
to randomly initialized bit strings for different problem sizes.
These distributions correspond to the data used
to compute the mean values and standard deviations
of the greedy (random) results shown in Fig.~\ref{fig:cost_convergence},
where they are summarized as error bars.
Since the final post-processing step of PCE
is also based on greedy local greedy search,
this baseline provides a reference for assessing
how informative the decoded binary state is
before post-processing.

Across all problem sizes,
the cost values obtained from state-vector simulations
are consistently lower than those obtained from QPU experiments.
Moreover, the simulation results remain below
the typical range of the greedy (random) distribution,
showing that the decoded bit strings obtained from PCE
provide systematically better starting points
for greedy refinement than random initializations.

For larger instances ($m = 2{,}772$ and $m = 10{,}296$),
the QPU results take values close to the mean
of the greedy (random) distribution,
indicating that the decoded QPU states become
close to effectively random initializations
at the level of post-processed cost.
For smaller and intermediate sizes, however,
the QPU results yield consistently lower costs
than the greedy (random) baseline,
demonstrating that the decoded bit strings
still retain useful problem-dependent structure
before post-processing.
These trends are consistent with the behavior observed
in Fig.~\ref{fig:cost_convergence}.

\subsection{Correlator Distributions}
\label{sec:correlator_appendix}
Figure~\ref{fig:correlator_6fig} presents the distributions of Pauli correlators
$\langle \Pi_i^{(k)} \rangle_{\bm \theta}$ across different problem sizes,
comparing state-vector simulations and QPU experiments.
These distributions are particularly useful
for interpreting the behavior of the QPU results
relative to the greedy (random) baseline in Fig.~\ref{fig:cost_convergence},
since they reveal how much problem-dependent structure
remains in the measured correlators before decoding and post-processing.
Each histogram is constructed from the full set of correlators
$\{\langle \Pi_i^{(k)} \rangle\}$, which take values in $[-1,1]$.
Since the distributions are strongly concentrated near zero,
we focus on the interval $[-0.1,0.1]$ to resolve their structure.

A clear size-dependent transition is observed in the simulation results.
For smaller problem sizes [Figs.~\ref{fig:correlator_6fig}(a)--(c)],
the correlator distributions exhibit discrete peak structures,
reflecting the finite number of variables
and the limited resolution of the representation.
In contrast, for larger problem sizes [Figs.~\ref{fig:correlator_6fig}(d)--(f)],
the distributions become effectively continuous.

In the large-scale regime, the simulation results further exhibit
a characteristic double-peak structure around zero,
with peaks symmetrically located at finite values of
$\langle \Pi_i^{(k)} \rangle$.
This behavior indicates that the correlator representation
develops a broader and more structured distribution,
which enables small continuous changes
to be transmitted more reliably
to the decoded binary variables.
By contrast, in small systems,
the correlator distributions remain sparse and effectively discrete,
so that continuous updates are less likely
to induce changes in the decoded binary state.

The QPU results show systematic deviations
from the ideal simulation behavior,
which can be attributed to finite sampling noise
and hardware imperfections.
For large problem sizes [Figs.~\ref{fig:correlator_6fig}(e) and (f)],
the QPU distributions closely resemble
a centered binomial distribution,
indicating that the correlators are broadly distributed around zero
with substantially reduced structural features.
This loss of structure is consistent
with the observation in Appendix~\ref{sec:cost_hist_appendix}
that the corresponding post-processed costs
approach the greedy (random) baseline.

For the intermediate case [Fig.~\ref{fig:correlator_6fig}(d)],
the simulation exhibits a broadened double-peak structure,
whereas the QPU result shows a predominantly single-peaked distribution
with slight broadening.
A comparison with the binomial reference suggests
that the central weight is reduced
and weak side structures persist,
which can be interpreted as a residual signature
of the simulation distribution.

For smaller problem sizes [Figs.~\ref{fig:correlator_6fig}(a)--(c)],
the QPU results exhibit broadened and dispersed peaks
relative to the discrete simulation distributions.
This broadening originates primarily
from finite-shot fluctuations (bounded by $1/\sqrt{N_{\mathrm{shot}}} \approx 0.016$),
which smear the discrete expectation values
while still preserving visible problem-dependent structure.

Overall, although quantitative discrepancies
between simulation and QPU results are evident,
the observed distributions can be consistently interpreted
in terms of three effects:
(i) regularization-induced concentration of correlators near zero,
(ii) finite-sampling fluctuations in expectation value estimation,
and (iii) hardware noise.
These observations support the interpretation presented in the main text,
namely that the optimization behavior is governed
by how effectively the correlator representation
can transmit continuous updates
to the decoded binary solution.

\section{Detailed Power Procurement Profiles}
\label{sec:appendix_profiles}

\begin{figure*}[t]
  \centering
  \includegraphics[width=0.32\textwidth]{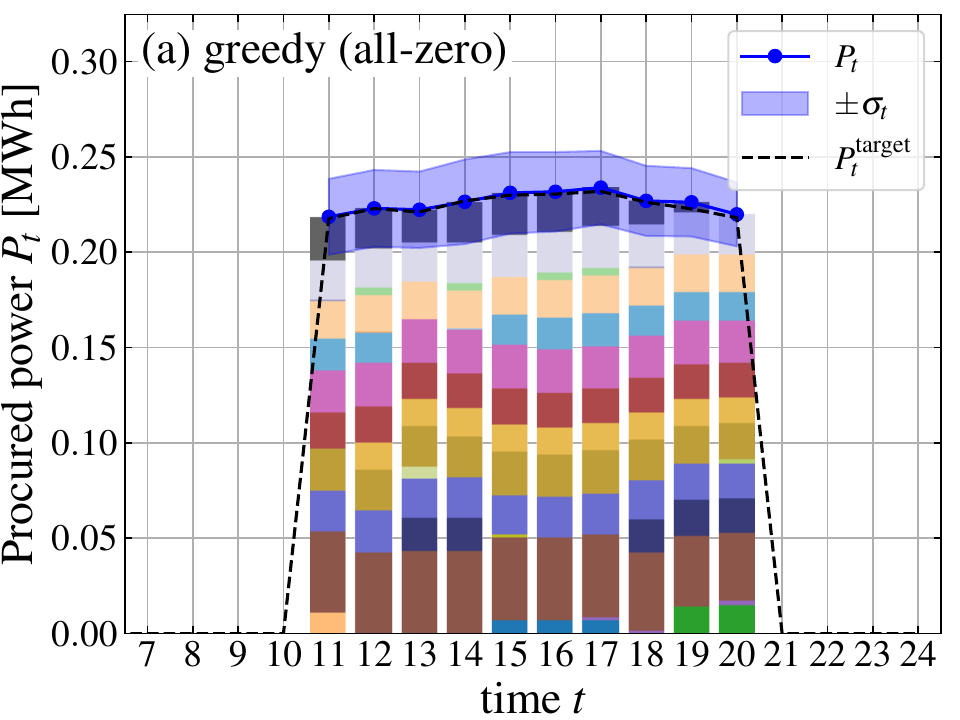}
  \hfill
  \includegraphics[width=0.32\textwidth]{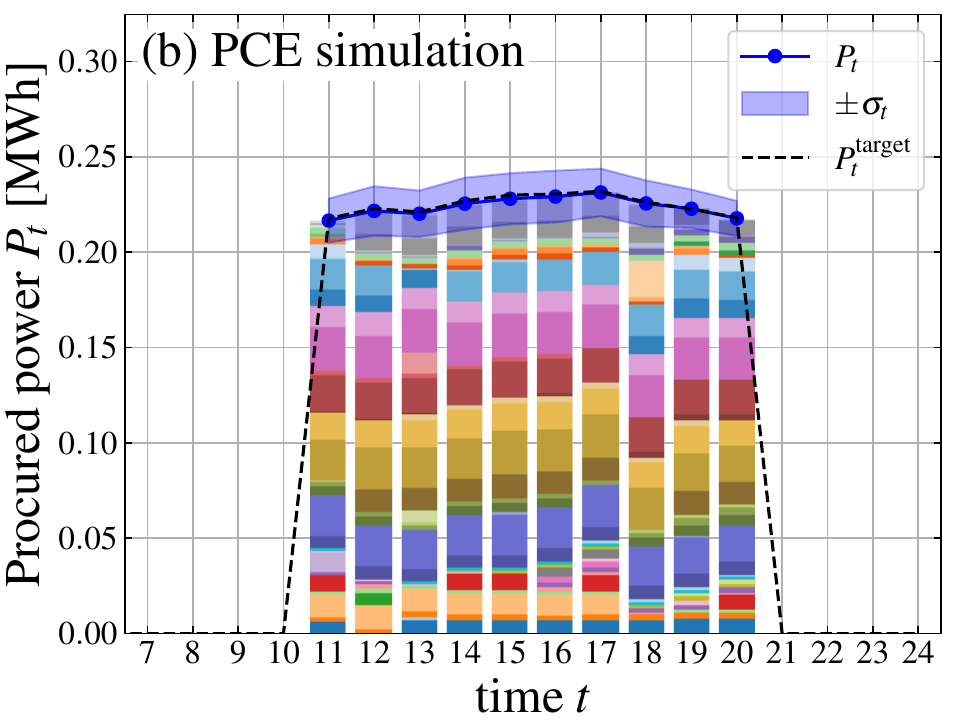}
  \hfill
  \includegraphics[width=0.32\textwidth]{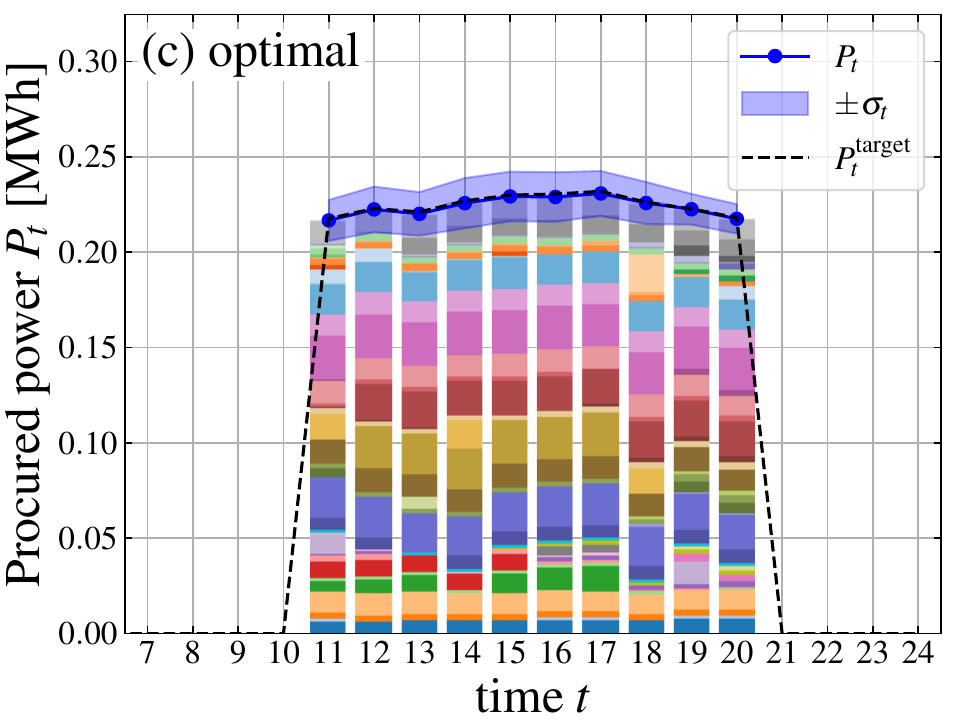}
  \caption{\label{fig:negawatt_std_stack_m60}
    Power procurement profiles for the time-resolved formulation (Model~2) with $m=60$ consumers.
  Solutions obtained by (a) the greedy method, (b) the Pauli correlation encoding (PCE)-based method,
  and (c) the optimal solution.
  In each panel, the target power $P^{\mathrm{target}}_t$ is indicated by a black dashed line,
  the expected procured power $P_t$ is shown as a solid line,
  and the shaded region represents the uncertainty range of $\pm \sigma_t$.
  The stacked area plot visualizes procured power from selected consumers, color-coded by consumer,
  where the top of the stack corresponds to $P_t$.
  Results are shown for hourly time steps from $t=11$ to $20$, each optimized independently.
  The optimal solution was obtained using the commercial solver Gurobi (version 12.0.3)~\cite{gurobi},
  with the MIP optimality tolerance (relative gap $10^{-4}$).
  }
\end{figure*}

\begin{figure*}[t]
  \centering
  \includegraphics[width=0.32\textwidth]{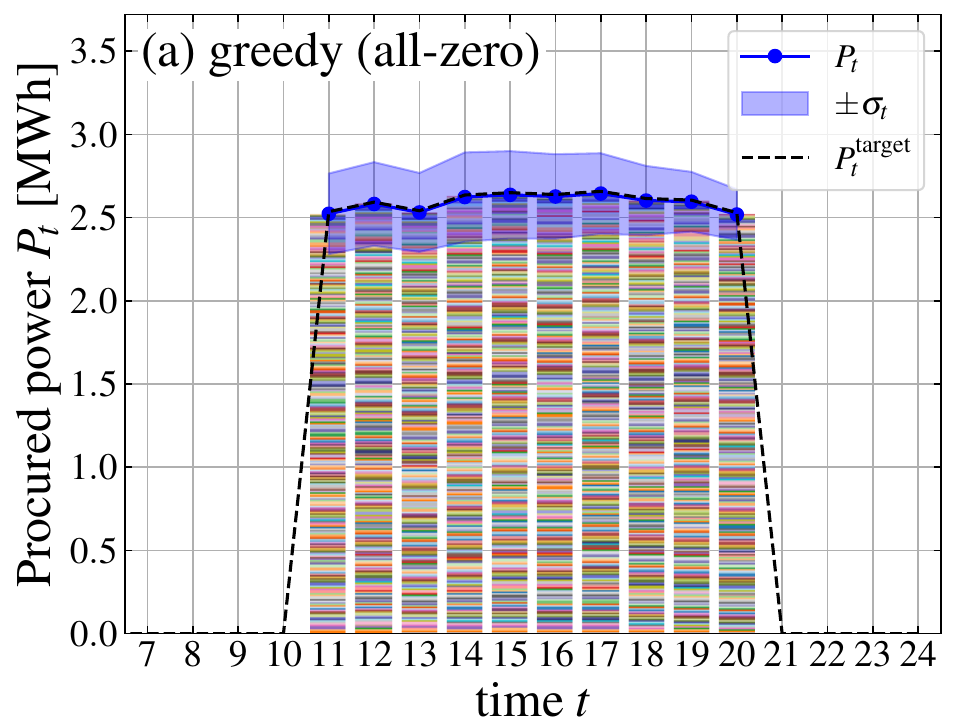}
  \hfill
  \includegraphics[width=0.32\textwidth]{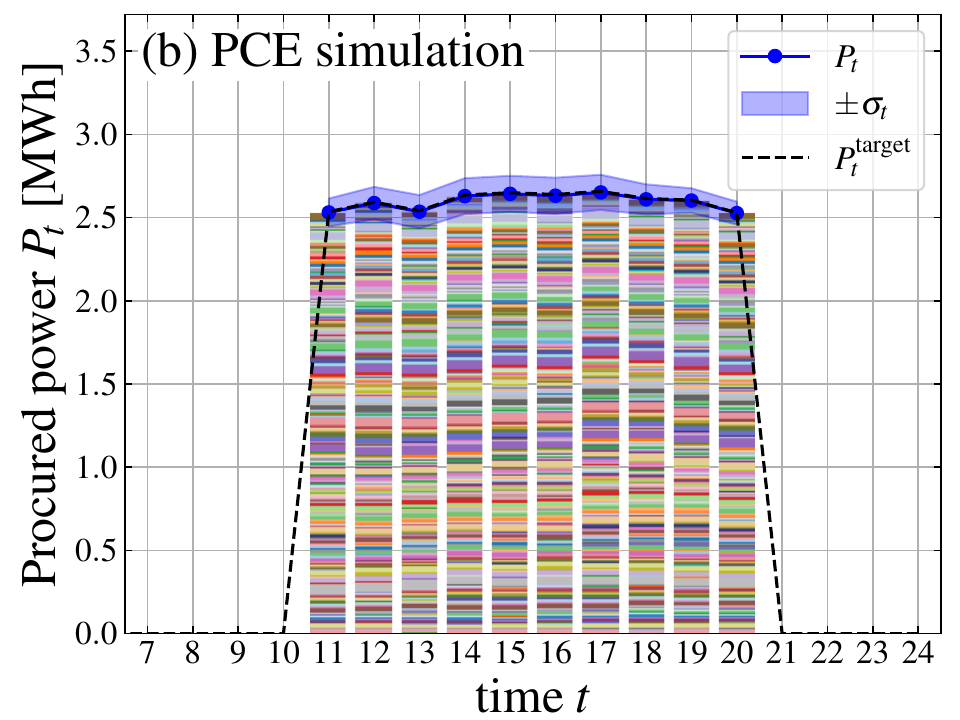}
  \hfill
  \includegraphics[width=0.32\textwidth]{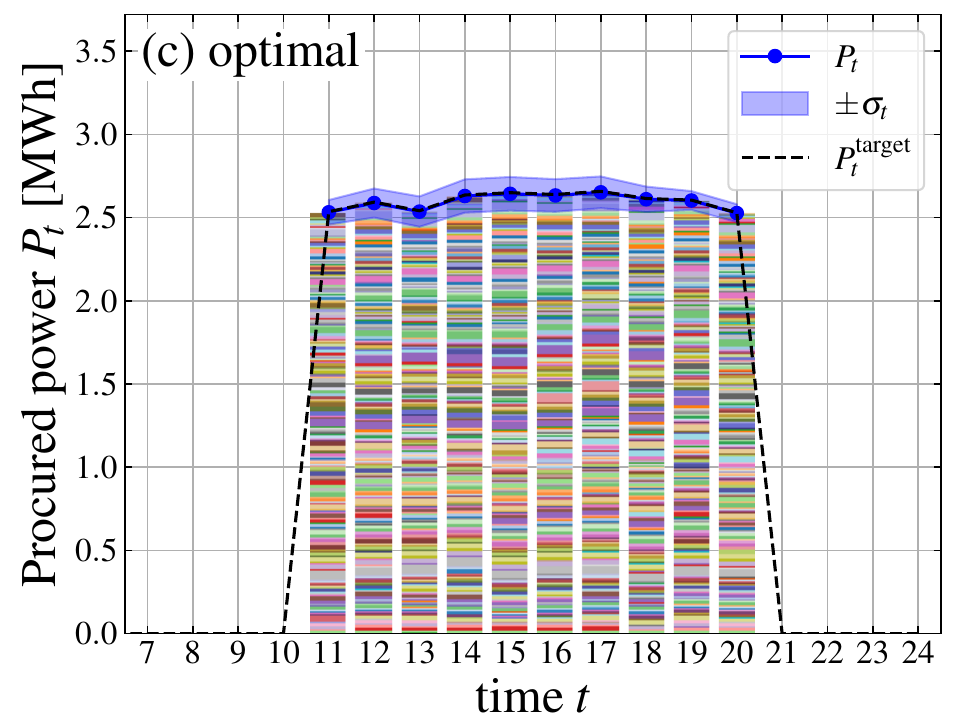}
  \caption{\label{fig:negawatt_std_stack_m756}
    Same as Fig.~\ref{fig:negawatt_std_stack_m60}, but for the instance with $m=756$ consumers.
  }
\end{figure*}

Figures~\ref{fig:negawatt_std_stack_m60} and \ref{fig:negawatt_std_stack_m756}
show detailed power procurement profiles for problem instances with $m=60$ and $m=756$ consumers, respectively.
These figures correspond to the accuracy and stability metrics summarized in
Fig.~\ref{fig:deviation_and_CV}, namely the absolute deviation
$| P_t - P^{\mathrm{target}}_t|$ and the coefficient of variation (CV) $\sigma_t/P_t$,
and provide a time-resolved visualization of how demand-side flexibility
from multiple consumers is aggregated to meet the target power in a stable manner.

For the smaller instance with $m=60$ consumers
(Fig.~\ref{fig:negawatt_std_stack_m60}),
the stacked representation reveals that the PCE-based solution
closely follows the optimal procurement pattern,
indicating that the method captures the underlying structure
of the optimal portfolio.

For the larger instance with $m=756$ consumers
(Fig.~\ref{fig:negawatt_std_stack_m756}),
individual contributions are visually aggregated.
Compared with the greedy baseline,
the PCE-based solution exhibits a narrower uncertainty band
and improved alignment with the procurement target,
consistent with the reduced deviation and CV
observed in Fig.~\ref{fig:deviation_and_CV}.

For the largest instance ($m=10{,}296$),
a similar stacked visualization is not shown,
as the large number of consumers leads to substantial visual crowding.
Instead, the aggregate performance is summarized
through the deviation and CV metrics in Fig.~\ref{fig:deviation_and_CV}.

\bibliographystyle{apsrev4-2}
\bibliography{references}

@CONTROL{apsrev42Control,
  author="00",
  editor="1",
  pages="1",
  title="0",
  year="0"
}

@article{Peruzzo2014,
	author = {Peruzzo, Alberto and McClean, Jarrod and Shadbolt, Peter and Yung, Man-Hong and Zhou, Xiao-Qi and Love, Peter J. and Aspuru-Guzik, Al{\'a}n and O'Brien, Jeremy L.},
	date = {2014/07/23},
	doi = {10.1038/ncomms5213},
	id = {Peruzzo2014},
	isbn = {2041-1723},
	journal = {Nature Communications},
	number = {1},
	pages = {4213},
	title = {A variational eigenvalue solver on a photonic quantum processor},
	url = {https://doi.org/10.1038/ncomms5213},
	volume = {5},
	year = {2014}}

@article{Cerezo2021,
	author = {Cerezo, M. and Arrasmith, Andrew and Babbush, Ryan and Benjamin, Simon C. and Endo, Suguru and Fujii, Keisuke and McClean, Jarrod R. and Mitarai, Kosuke and Yuan, Xiao and Cincio, Lukasz and Coles, Patrick J.},
	date = {2021/09/01},
	doi = {10.1038/s42254-021-00348-9},
	id = {Cerezo2021},
	isbn = {2522-5820},
	journal = {Nature Reviews Physics},
	number = {9},
	pages = {625--644},
	title = {Variational quantum algorithms},
	url = {https://doi.org/10.1038/s42254-021-00348-9},
	volume = {3},
	year = {2021}}

@article{Biamonte2017,
	author = {Biamonte, Jacob and Wittek, Peter and Pancotti, Nicola and Rebentrost, Patrick and Wiebe, Nathan and Lloyd, Seth},
	date = {2017/09/01},
	doi = {10.1038/nature23474},
	id = {Biamonte2017},
	isbn = {1476-4687},
	journal = {Nature},
	number = {7671},
	pages = {195--202},
	title = {Quantum machine learning},
	url = {https://doi.org/10.1038/nature23474},
	volume = {549},
	year = {2017}}

@article{Mitarai2018,
  title = {Quantum circuit learning},
  author = {Mitarai, K. and Negoro, M. and Kitagawa, M. and Fujii, K.},
  journal = {Phys. Rev. A},
  volume = {98},
  issue = {3},
  pages = {032309},
  numpages = {6},
  year = {2018},
  month = {Sep},
  publisher = {American Physical Society},
  doi = {10.1103/PhysRevA.98.032309},
  url = {https://link.aps.org/doi/10.1103/PhysRevA.98.032309}
}

@article{Tan2021,
  doi = {10.22331/q-2021-05-04-454},
  url = {https://doi.org/10.22331/q-2021-05-04-454},
  title = {Qubit-efficient encoding schemes for binary optimisation problems},
  author = {Tan, Benjamin and Lemonde, Marc-Antoine and Thanasilp, Supanut and Tangpanitanon, Jirawat and Angelakis, Dimitris G.},
  journal = {{Quantum}},
  issn = {2521-327X},
  publisher = {{Verein zur F{\"{o}}rderung des Open Access Publizierens in den Quantenwissenschaften}},
  volume = {5},
  pages = {454},
  month = may,
  year = {2021}
}

@article{Glos2022,
	author = {Glos, Adam and Krawiec, Aleksandra and Zimbor{\'a}s, Zolt{\'a}n},
	date = {2022/04/19},
	doi = {10.1038/s41534-022-00546-y},
	id = {Glos2022},
	isbn = {2056-6387},
	journal = {npj Quantum Information},
	number = {1},
	pages = {39},
	title = {Space-efficient binary optimization for variational quantum computing},
	url = {https://doi.org/10.1038/s41534-022-00546-y},
	volume = {8},
	year = {2022}}

@misc{Teramoto2023,
      title={Quantum-Relaxation Based Optimization Algorithms: Theoretical Extensions}, 
      author={Kosei Teramoto and Rudy Raymond and Eyuri Wakakuwa and Hiroshi Imai},
      year={2023},
      eprint={2302.09481},
      archivePrefix={arXiv},
      primaryClass={quant-ph},
      url={https://arxiv.org/abs/2302.09481}, 
}

@article{He2025,
	author = {He, Zichang and Raymond, Rudy and Shaydulin, Ruslan and Pistoia, Marco},
	date = {2025/08/09},
	doi = {10.1038/s41598-025-13543-w},
	id = {He2025},
	isbn = {2045-2322},
	journal = {Scientific Reports},
	number = {1},
	pages = {29191},
	title = {Non-variational quantum random access optimization with alternating operator ansatz},
	url = {https://doi.org/10.1038/s41598-025-13543-w},
	volume = {15},
	year = {2025}}

@INPROCEEDINGS{Sharma2024,
  author={Sharma, Monit and Jin, Yan and Lau, Hoong Chuin and Raymond, Rudy},
  booktitle={2024 IEEE International Conference on Quantum Computing and Engineering (QCE)}, 
  title={Quantum Relaxation for Solving Multiple Knapsack Problems}, 
  year={2024},
  volume={01},
  number={},
  pages={692-698},
  doi={10.1109/QCE60285.2024.00086}}

@INPROCEEDINGS{Matsuyama2024,
  author={Matsuyama, Hiromichi and Huang, Wei-hao and Nishimura, Kohji and Yamashiro, Yu},
  booktitle={2024 IEEE International Conference on Quantum Computing and Engineering (QCE)}, 
  title={Efficient Internal Strategies in Quantum Relaxation Based Branch-and-Bound}, 
  year={2024},
  volume={01},
  number={},
  pages={470-480},
  doi={10.1109/QCE60285.2024.00062}}

@article{Albadi2008,
	author = {M.H. Albadi and E.F. El-Saadany},
	doi = {https://doi.org/10.1016/j.epsr.2008.04.002},
	issn = {0378-7796},
	journal = {Electric Power Systems Research},
	number = {11},
	pages = {1989-1996},
	title = {A summary of demand response in electricity markets},
	url = {https://www.sciencedirect.com/science/article/pii/S0378779608001272},
	volume = {78},
	year = {2008}}

@article{Palensky2011,
	author = {Palensky, Peter and Dietrich, Dietmar},
	doi = {10.1109/TII.2011.2158841},
	journal = {IEEE Transactions on Industrial Informatics},
	number = {3},
	pages = {381-388},
	title = {Demand Side Management: Demand Response, Intelligent Energy Systems, and Smart Loads},
	volume = {7},
	year = {2011}}

@article{Sciorilli2025PCE,
  author       = {Marco Sciorilli and Lucas Borges and Taylor L. Patti and Diego Garc\'{\i}a-Mart\'{\i}n and Giancarlo Camilo and Anima Anandkumar and Leandro Aolita},
  title        = {Towards large-scale quantum optimization solvers with few qubits},
  journal      = {Nature Communications},
  volume       = {16},
  pages        = {476},
  year         = {2025},
  doi          = {10.1038/s41467-024-55346-z},
  url          = {https://www.nature.com/articles/s41467-024-55346-z},
}

@article{Ajagekar2019,
	author = {Akshay Ajagekar and Fengqi You},
	doi = {10.1016/j.energy.2019.04.186},
	journal = {Energy},
	month = {jul},
	pages = {76--89},
	publisher = {Elsevier {BV}},
	title = {Quantum computing for energy systems optimization: Challenges and opportunities},
	url = {https://doi.org/10.1016/j.energy.2019.04.186},
	volume = {179},
	year = 2019}

@techreport{CRIEPI_C18005,
  author       = {Tsurumi, Tsuyoya and Syoji, Kenichi},
  title        = {A design method for electric power demand portfolio and its basic investigation -- Proposition of a method for a resource aggregator to procure negawatt effectively --},
  institution  = {Central Research Institute of Electric Power Industry},
  address      = {Tokyo, Japan},
  number       = {C18005},
  note         = {CRIEPI Technical Report},
  year         = {2019},
  url          = {https://criepi.denken.or.jp/hokokusho/pb/reportDetail?reportNoUkCode=C18005}
}

@article{Faia2021,
	author = {Ricardo Faia and Tiago Pinto and Zita Vale and Juan Manuel Corchado},
	doi = {https://doi.org/10.1016/j.ijepes.2020.106739},
	issn = {0142-0615},
	journal = {International Journal of Electrical Power and Energy Systems},
	pages = {106739},
	title = {Portfolio optimization of electricity markets participation using forecasting error in risk formulation},
	url = {https://www.sciencedirect.com/science/article/pii/S0142061520342848},
	volume = {129},
	year = {2021}}

@misc{Farhi2014QAOA,
      title={A Quantum Approximate Optimization Algorithm}, 
      author={Edward Farhi and Jeffrey Goldstone and Sam Gutmann},
      year={2014},
      eprint={1411.4028},
      archivePrefix={arXiv},
      primaryClass={quant-ph},
      url={https://arxiv.org/abs/1411.4028}, 
}

@article{Hadfield2019,
	article-number = {34},
	author = {Hadfield, Stuart and Wang, Zhihui and O'Gorman, Bryan and Rieffel, Eleanor G. and Venturelli, Davide and Biswas, Rupak},
	doi = {10.3390/a12020034},
	issn = {1999-4893},
	journal = {Algorithms},
	number = {2},
	title = {From the Quantum Approximate Optimization Algorithm to a Quantum Alternating Operator Ansatz},
	url = {https://www.mdpi.com/1999-4893/12/2/34},
	volume = {12},
	year = {2019}}

@inproceedings{YoshiokaFQAOA2024,
        author = {Yoshioka, Takuya and Sasada, Keita and Nakano, Yuichiro and Fujii, Keisuke},
        booktitle = {2024 IEEE International Conference on Quantum Computing and Engineering (QCE)},
        doi = {10.1109/QCE60285.2024.00061},
        pages = {469-475},
        title = {Electric Power Demand Portfolio Optimization by Fermionic QAOA with Self-Consistent Local Field Modulation},
        volume = {01},
        year = {2024}}

@inproceedings{YoshiokaQCE2026,
  author    = {Yoshioka, Takuya and Sasada, Keita and Usuki, Riku and Fujii, Keisuke},
  title     = {Performance Analysis of Pauli Correlation Encoding for Electric Power Portfolio Optimization under Varying Procurement Levels},
  booktitle = {2026 IEEE International Conference on Quantum Computing and Engineering (QCE)},
  year       = {2026},
  note       = {Accepted for publication}
}

@misc{AmazonBraket,
  author       = {{Amazon Web Services}},
  title        = {Amazon Braket Documentation},
  howpublished = {\url{https://docs.aws.amazon.com/braket/}},
  year         = {2026}
}

@misc{gurobi,
  author       = {{Gurobi Optimization, LLC}},
  title        = {Gurobi Optimizer Reference Manual},
  howpublished = {\url{https://www.gurobi.com}},
  year         = {2026}
}

@misc{ems_opendata,
  author       = {{Sustainable open Innovation Initiative}},
  title        = {Energy Management System},
  howpublished = {\url{https://www.ems-opendata.jp}},
  year         = {2026}
}

@article{qulacs,
  doi = {10.22331/q-2021-10-06-559},
  url = {https://doi.org/10.22331/q-2021-10-06-559},
  title = {Qulacs: a fast and versatile quantum circuit simulator for research purpose},
  author = {Suzuki, Yasunari and Kawase, Yoshiaki and Masumura, Yuya and Hiraga, Yuria and Nakadai, Masahiro and Chen, Jiabao and Nakanishi, Ken M. and Mitarai, Kosuke and Imai, Ryosuke and Tamiya, Shiro and Yamamoto, Takahiro and Yan, Tennin and Kawakubo, Toru and Nakagawa, Yuya O. and Ibe, Yohei and Zhang, Youyuan and Yamashita, Hirotsugu and Yoshimura, Hikaru and Hayashi, Akihiro and Fujii, Keisuke},
  journal = {{Quantum}},
  issn = {2521-327X},
  publisher = {{Verein zur F{\"{o}}rderung des Open Access Publizierens in den Quantenwissenschaften}},
  volume = {5},
  pages = {559},
  month = oct,
  year = {2021}
}

@article{Soloviev2026Portfolio,
  author  = {Soloviev, Vicente P. and Krompiec, Michal},
  title   = {Large-scale portfolio optimization using Pauli correlation encoding},
  journal = {Scientific Reports},
  volume  = {16},
  pages   = {25158},
  year    = {2026},
  doi     = {10.1038/s41598-026-54244-2},
  url     = {https://www.nature.com/articles/s41598-026-54244-2}
}

@misc{Padinmartinez2026,
      title={Pauli Correlation Encoding for Budget-Constrained Optimization}, 
      author={Jacobo Padín-Martínez and Vicente P. Soloviev and Alejandro Borrallo-Rentero and Antón Rodríguez-Otero and Raquel Alfonso-Rodríguez and Michal Krompiec},
      year={2026},
      eprint={2602.17479},
      archivePrefix={arXiv},
      primaryClass={quant-ph},
      url={https://arxiv.org/abs/2602.17479}, 
}

@misc{Sciorilli2025LABS,
  author = {Sciorilli, M. and Camilo, G. and Maciel, T. O. and Canabarro, A.},
  title  = {A competitive NISQ and qubit-efficient solver for the LABS problem},
  year   = {2025},
  eprint = {2506.17391},
  archivePrefix = {arXiv},
  primaryClass  = {quant-ph}
}

@misc{Sharma2025Comparative,
  title={A Comparative Study of Quantum Optimization Techniques for Solving Combinatorial Optimization Benchmark Problems},
  author={Monit Sharma and Hoong Chuin Lau},
  year={2025},
  eprint={2503.12121},
  archivePrefix={arXiv},
  primaryClass={quant-ph}
}

\end{document}